\documentclass[review]{elsarticle}

\UseRawInputEncoding
\usepackage{lineno,hyperref}
\usepackage{xcolor}
\usepackage{amsmath}
\usepackage{physics}
\usepackage{tabularx}
\usepackage{multicol}
\usepackage{siunitx}
\usepackage{bm}
\usepackage{nomencl}
\usepackage{changepage}

\newcolumntype{C}{>{\centering\arraybackslash}X}
\modulolinenumbers[5]

\makenomenclature
\renewcommand\nomgroup[1]{%
  \item[\bfseries
  \ifstrequal{#1}{L}{List of notations}{%
  \ifstrequal{#1}{G}{Greek symbols}{%
  \ifstrequal{#1}{A}{Abbreviatures}{}}}%
]}

\renewcommand*\nompreamble{\begin{multicols}{2} \footnotesize}
\renewcommand*\nompostamble{\end{multicols}}

\journal{International Journal of Multiphase Flow}

\graphicspath{{figures/}}

\begin{document}

\begin{frontmatter}

\title{A computational methodology to account for the liquid film thickness evolution in Direct Numerical Simulation of prefilming airblast atomization}

\author{R. Payri\textsuperscript{1}, F.J. Salvador\textsuperscript{1}}
\author{M. Carreres\textsuperscript{1}\corref{mycorrespondingauthor}}
\author{C. Moreno-Montagud\textsuperscript{1}}
%\fntext[myfootnote]{CMT-Motores T\'ermicos, Universitat Polit\`ecnica de Val\`encia, Camino de Vera s/n, E-46022 Spain}

\author[mymainaddress]{CMT-Motores T\'ermicos, Universitat Polit\`ecnica de Val\`encia, Camino de Vera s/n, E-46022 Spain}

%\author[mysecondaryaddress]{Global Customer Service}
\cortext[mycorrespondingauthor]{Corresponding author}
\ead{marcarta@mot.upv.es}

\begin{abstract}
Prefilming airblast atomization is becoming widely used in current aero engines. Fundamental studies on the actual annular configuration of airblast atomizers are difficult to realize. For this reason, researchers have also focused on planar configurations. In this regard, the Karlsruhe Institute of Technology (KIT) developed a test rig to conduct experimental activities, conforming a large database with results for different working fluids and operating conditions. Such data allow two-phase flow modellers to validate their calculations concerning primary atomization on these devices. The present investigation proposes a Direct Numerical Simulation (DNS) study on the KIT planar configuration through the Volume of Fluid (VOF) method within the PARIS Simulator code. The novelty compared to DNS works reported in the literature resides in the use of a boundary condition that allows accounting not only for the gas inflow turbulence but also for the spatio-temporal evolution of the liquid film thickness at the DNS inlet and its related effect on turbulence. The proposed methodology requires computing precursor single-phase and two-phase flow Large-Eddy Simulations on the prefilmer flow. Results are compared to DNS that only account for a constant (both timewise and spanwise) liquid film thickness at the domain inlet, validating the full methodology workflow. The proposed methodology is shown to improve the qualitative description of the atomization mechanism, as the different stages of breakup (liquid accumulation behind the prefilmer edge, bag formation, bag breakup, ligament formation and ligament breakup) coexist spanwise for a given temporal snapshot. This implies a more continuous atomization than the one predicted by the constant film thickness case, which showed the same breakup stage to be present along the prefilmer span for a given instant and led to a more discretized set of atomization events. The proposed workflow allows quantifying the influence of the liquid film flow evolution above the prefilmer surface on primary breakup frequency and relevant atomization features.
\end{abstract}


\end{frontmatter}

% Nomenclature %%%%%%%%%%%%%%%%%%%%%%%

% LIST OF NOTATIONS

\nomenclature[L]{$b$}{Prefilmer span}
\nomenclature[L]{$C_{\alpha}$}{Interface artificial compression coefficient}
\nomenclature[L]{$\vb{f_{\sigma}}$}{Surface tension force vector}
\nomenclature[L]{$d_V$}{Droplet volumetric diameter}
\nomenclature[L]{$f_{bu}$}{Mean breakup frequency}
\nomenclature[L]{$f_{main}$}{Frequency among main bag breakup events}
\nomenclature[L]{$f_{film}$}{Liquid film wave frequency}
\nomenclature[L]{$g_c$}{Cell grading coefficient}
\nomenclature[L]{$h_c$}{Prefilmer channel height}
\nomenclature[L]{$h_l$}{Liquid film thickness}
\nomenclature[L]{$h_p$}{Prefilmer edge thickness}
\nomenclature[L]{$L_{bu}$}{Mean breakup length}
\nomenclature[L]{$L_f$}{Effective prefilming length}
\nomenclature[L]{$L_p$}{Prefilmer length}
\nomenclature[L]{$L_{str}$}{Ligament length}
\nomenclature[L]{$L_x$}{DNS domain length}
\nomenclature[L]{$L_y$}{DNS domain height}
\nomenclature[L]{$L_z$}{DNS domain width}
\nomenclature[L]{$l_{\eta}$}{Kolmogorov length scale}
\nomenclature[L]{$M$}{Momentum flux ratio}
%\nomenclature[L]{$m$}{Rosin-Rammler scale parameter}
\nomenclature[L]{$\vb{n}$}{Interface normal vector}
\nomenclature[L]{$Oh$}{Ohnesorge number}
\nomenclature[L]{$p$}{Pressure}
%\nomenclature[L]{$q$}{Rosin-Rammler shape parameter}
\nomenclature[L]{$Re$}{Reynolds number}
\nomenclature[L]{$Re_{\tau}$}{Frictional Reynolds number}
\nomenclature[L]{$T$}{Temperature}
\nomenclature[L]{$t$}{Time}
\nomenclature[L]{$\vb{u} = (u, v, w)$}{Velocity vector}
\nomenclature[L]{$\vb{u_c}$}{Compression velocity vector}
\nomenclature[L]{$\vb{u_{def}}$}{Film deformation velocity}
\nomenclature[L]{$\vb{u_{lig}}$}{Mean ligament tip velocity vector}
\nomenclature[L]{$\vb{u_{str}}$}{Ligament tip velocity vector}
\nomenclature[L]{$u_m$}{Average streamwise velocity at the DNS inlet}
\nomenclature[L]{$V$}{Volume}
\nomenclature[L]{$\dot{V}$}{Volumetric flow rate}
\nomenclature[L]{$We$}{Weber number}
\nomenclature[L]{$\vb{x} = (x, y , z)$}{Position vector}
\nomenclature[L]{$y^+$}{Non-dimensional boundary layer distance}

%GREEK SYMBOLS

\nomenclature[G]{$\alpha$}{Liquid volume fraction}
\nomenclature[G]{$\delta$}{Boundary layer thickness}
\nomenclature[G]{$\delta_S$}{Dirac delta function}
\nomenclature[G]{$\kappa$}{Interface curvature}
\nomenclature[G]{$\lambda$}{Liquid film wavelength}
\nomenclature[G]{$\mu$}{Dynamic viscosity}
\nomenclature[G]{$\nu$}{Kinematic viscosity}
\nomenclature[G]{$\rho$}{Density}
\nomenclature[G]{$\sigma$}{Surface tension}
\nomenclature[G]{$\theta$}{Liquid-wall contact angle}

%ABBREVATIONS

\nomenclature[A]{CFD}{Computational Fluid Dynamics}
\nomenclature[A]{CFL}{Courant-Friedrichs-Lewy number}
\nomenclature[A]{CLSMOF}{Coupled Level-Set Moment of Fluid}
\nomenclature[A]{CLSVOF}{Coupled Level-Set Volume of Fluid}
\nomenclature[A]{CSF}{Continuum Surface Force}
\nomenclature[A]{DMD}{Dynamic Mode Decomposition}
\nomenclature[A]{DNS}{Direct Numerical Simulation}
\nomenclature[A]{eDNS}{embedded Direct Numerical Simulation}
\nomenclature[A]{ELSA}{Eulerian-Lagrangian Spray Atomization}
\nomenclature[A]{FCT}{Flux-Corrected Transport}
\nomenclature[A]{GAMG}{Geometric Agglomerated Algebraic Multi-Grid}
\nomenclature[A]{HF}{Height Function}
\nomenclature[A]{ICM}{Interface Capturing Methods}
\nomenclature[A]{KIT}{Karslruhe Institute of Technology}
\nomenclature[A]{LES}{Large-Eddy Simulation}
\nomenclature[A]{LDA}{Laser Doppler Anemometry}
\nomenclature[A]{MULES}{Multidimensional Universal Limiter for Explicit Solution}
\nomenclature[A]{PAMELA}{Primary Atomization Model for prEfilming airbLAst injectors}
\nomenclature[A]{PARIS}{PArallel, Robust, Interface Simulator}
\nomenclature[A]{PDA}{Phase Doppler Anemometry}
\nomenclature[A]{PDF}{Probability Density Function}
\nomenclature[A]{PISO}{Pressure-Implicit with Splitting of Operators}
\nomenclature[A]{PLIC}{Piecewise Linear Interface Calculation}
\nomenclature[A]{POD}{Proper Orthogonal Decomposition}
\nomenclature[A]{RMS}{Root Mean Square}
\nomenclature[A]{SGS}{Sub-grid Scale}
\nomenclature[A]{SMD}{Sauter Mean Diameter}
\nomenclature[A]{SPH}{Smoothed Particle Hydrodynamics}
\nomenclature[A]{TE}{Trailing Edge}
\nomenclature[A]{VOF}{Volume of Fluid}
\nomenclature[A]{VTK}{The Visualization Toolkit}
\nomenclature[A]{WALE}{Wall Adapting Local Eddy-viscosity}

%%%%%%%%%%%%%%%%%%%%%%%%%%%%%%%%%%%%%%

\printnomenclature

%\linenumbers

\section{Introduction} \label{sec:introduction}

The aviation industry produces a not negligible part of the global pollutant emissions, due to its important growth in the last decades \cite{ArnaldoValdes2019}. This problem gets worse because combustion residues at high altitudes of the troposphere are more harmful to the environment than they are at sea level \cite{Sovde2014}. In order to lower the emissions, new engine concepts and combustion technologies are required. In this context, lean combustion emerges to improve fuel consumption and emissions, but requiring high control of the process to avoid flameout \cite{Liu2017a}. For this reason, the research community has increased their interest in the atomization process, as controlling combustion in aero engines highly depends on this phenomenon.

Some of the most commonly used atomizers in the aviation industry are pressure-swirl atomizers and prefilming airblast atomizers \cite{Lefebvre1988}. In the former, liquid fuel is brought into the environment and atomized by means of a pressure difference. In the latter, in turn, the fuel is first deposited onto a prefilmer and then driven by a surrounding high-velocity airstream towards an atomizing edge. When the liquid reaches this edge, the film is destabilized, it breaks up into ligaments and subsequently disintegrates into droplets. This study focuses on this last type of atomizers.

Some groups have performed experimental studies on actual annular configurations of airblast atomizers. Jasuja \cite{Jasuja2006} compared low- and high-shear design philosophies over a wide range of ambient air densities and kerosene flow rates at an ambient air temperature using Phase Doppler Interferometry and laser sheet imaging. Matsuura et al. \cite{Matsuura2008} investigated the effects of $p_a$ on SMD for a counter double-swirl high-shear-type fuel injector at different air pressure drops, kerosene flow rates and AFR than Jasuja \cite{Jasuja2006}, using Phase Doppler Anemometry and laser sheet Mie scattering visualization. Gepperth et al. \cite{gepperth2014primary} explored the coupling of the film flow and primary breakup process for an industrial prefilming airblast nozzle through high speed shadowgraphy. Two advanced mathematical procedures were used for the first time to analyze the high resolution recordings of such atomizers: proper orthogonal decomposition (POD) and dynamic mode decomposition (DMD). Results revealed an independence between the extracted frequency of the precessing vortex core, film waves and atomization events for this range of operating conditions. In addition, a linear behavior of the film flow dynamics was observed from the DMD analysis.

Nevertheless, the annular configuration of the airblast atomizer poses a multi-scale nature problem and presents some geometrical complexity, importantly hindering its study. In this sense, many researchers have focused their efforts on planar configurations, which are much simpler but allow extrapolating conclusions to the actual annular configurations due to an analogy by similarity, as pointed out by Berthoumieu and Lavergne \cite{Berthoumieu2001} and confirmed by Holz et al. \cite{Holz2016}.

\begin{figure}[htbp]
	\centering
	\includegraphics[width=0.5\textwidth]{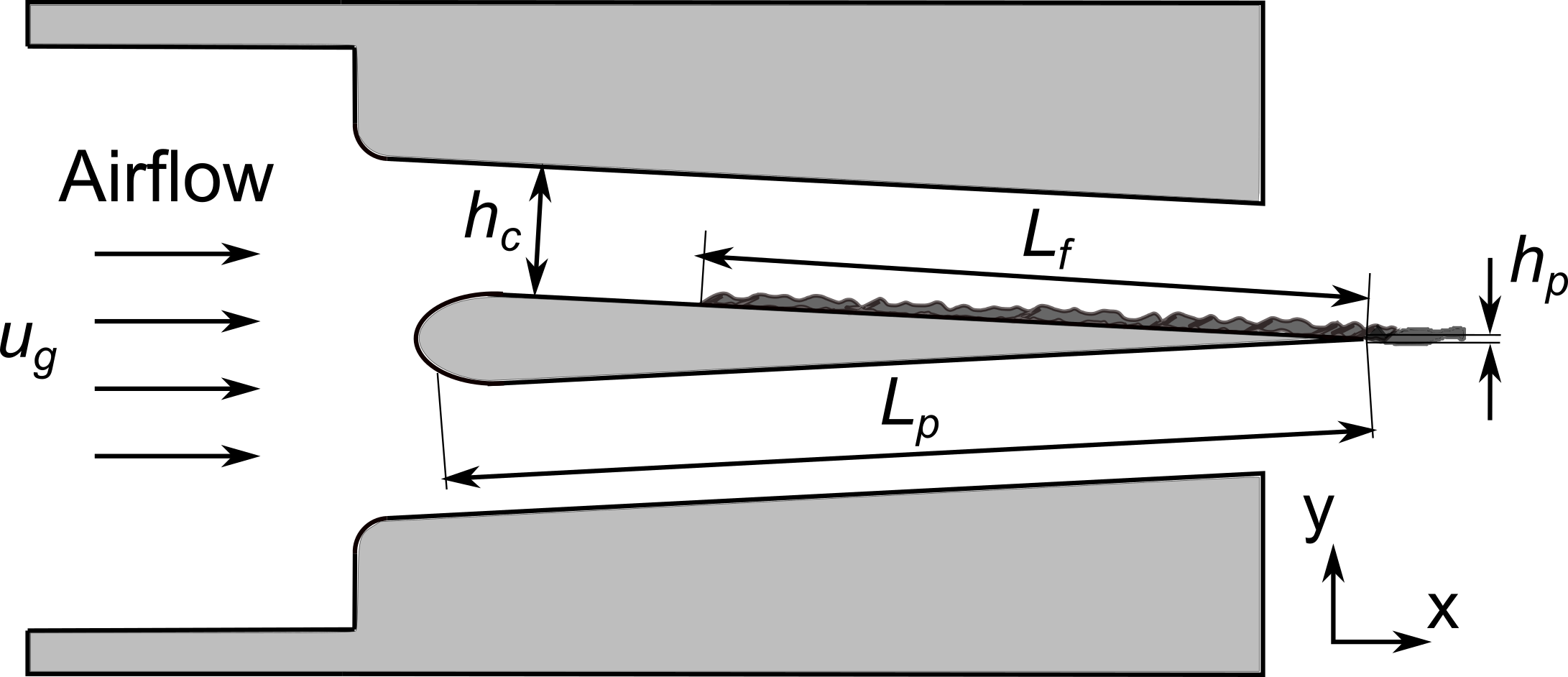}
	\centering
	\caption{Sketch of the KIT-ITS test rig cross-section.}
	\label{fig:KITTestRig}
\end{figure}

Focusing on the planar configuration, several test rigs have been developed to study the primary breakup in airblast atomizers, generally using an airfoil-shaped solid section as a prefilmer. Inamura et al. \cite{Inamura2012, Inamura2019} used this sort of test rig with water as a working fluid under different operating conditions, establishing the primary breakup mechanism of these devices. The test rig in KIT-ITS \cite{Gepperth2010}, in turn, stands out as it has been used to carry out tests with diverse operating conditions and fuels through a variety of techniques, providing a huge experimental database. In this particular model, the airflow is divided into two ducts by the mentioned airfoil-shaped solid section (as shown in Figure \ref{fig:KITTestRig}) and the fuel is injected through some holes equally distributed spanwise close to the leading edge of the upper surface. The fuel then develops as a planar liquid film reaching the atomizing edge.

% EXPERIMENTOS KIT
A first set of experimental studies reported in the KIT-ITS test rig were conducted by Gepperth et al. \cite{Gepperth2010,Gepperth2012}. For a varied range of operating conditions (fluids, flow rates, prefilmer geometrical dimensions, ...), the influence of many parameters in primary atomization has been studied at ambient pressure and temperature using shadowgraphy, PDA measurements and ligament tracking. The results showed a strong dependence of the mean droplet diameter on mean air velocity and atomizing edge thickness, these being the dominant parameters. Prefilming length, liquid physical properties and liquid volume flow rate had a weaker effect, but still affected the ligament formation process. From these results, correlations were derived to predict the breakup frequency, Sauter Mean Diameter (SMD) and mean droplet velocities in the primary atomization region. Unlike previously existing models, their correlation did not require knowledge of film properties, which are difficult to estimate.

In addition to those steady air flow experiments, Chaussonnet et al. \cite{Chaussonnet2017,Chaussonnet2018} carried out similar tests at ambient temperature and pressure but with fluctuating air flow instead, again using shadowgraphy and LDA velocimetry. A pulsating device generated velocity fluctuations of certain amplitude and frequency, corresponding to usual observed magnitudes in real combustion chambers. Results reported a significant influence on SMD up to a certain frequency, that is, a non-linear transfer function with a low-pass filter type of behaviour.

Results were later extended to higher ambient pressures through new experiments by Chaussonet et al. \cite{Chaussonnet2020}. The aerodynamic stress $\rho_g u_g^2$ was used instead of the ambient pressure as a more appropriate parameter to characterize prefilming airblast breakup, and the domain was virtually split into several atomizing cells with single ligaments. Besides, two new characteristic lengths were proposed: one based on atomizing cell streamwise surface, related to the air velocity; and another one related to film loading, which showed a correlation with the SMD. This is in line with the idea from all previous works that liquid accumulation is key to determine the primary spray characteristics. Many correlations from the literature were compared with these experiments, most of them underestimating SMD in their predictions. This emphasizes the need of calibrating models using experimental techniques like shadowgraphy \cite{Payri2020}.

% SIMULATION LES + PAMELA
As experimental studies are expensive and involve specialized facilities and resources, many researchers have delved into Computational Fluid Dynamics (CFD) to model primary atomization in pursuit of accuracy at lower costs. Chaussonet et al. \cite{Chaussonnet2016} studied the prefilming airblast atomizer geometry through Large Eddy Simulations (LES) to compare with Gepperth \cite{gepperth2014primary,Gepperth2012} experimental results. They developed the Primary Atomization Model for prEfilming airbLAst injectors (PAMELA) and implemented it into the LES solver AVBP. Considering the experimental observations, the droplet size distribution followed a Rosin-Rammler function with both scale and shape parameters depending on flow properties. The PAMELA model links these parameters with two Weber numbers based on the thickness of the atomizing edge and the boundary layer on the prefilmer ($We_{h_p}$ and $We_{\delta}$) and an extra function of $h_p$, obtaining five constants that are independent from the flow conditions. After calibration of these constants, they were able to predict mean and RMS velocity profiles, capture the vortex shedding downstream of the atomizing edge and the drop size Probability Density Function (PDF) of the spray with moderate computational effort. Palanti et al. \cite{Palanti2021} also carried out LES of the study case, coupling Interface Capturing Methods (ICM) with the Eulerian–Lagrangian Spray Atomization (ELSA) approach with the OpenFOAM open source code. Their objective was to catch the main features of the breakup mechanisms with a low-intensity computational strategy and a novel postprocessing technique, relating curvature of the liquid interface to the PDF function.

%SIMULATION DNS
To get a better understanding of the atomization process, other authors opted for Direct Numerical Simulations (DNS) as they can provide detailed information hardly achievable by current optical techniques. Nevertheless, accurate DNS of a full prefilming airblast atomizer geometry are unreasonable with current computational resources, even assuming the planar configuration simplification. To overcome this limitation, these authors relegated the domain to the last part of the prefilmer and the primary breakup zone.

%BOUNDARIES + LES PRECURSORAS
Mukundan et al. \cite{Mukundan2019a} and Braun et al. \cite{Braun2019} used constant velocity profiles for both the liquid and gas phases at the inlet boundary conditions, gaining some insight on the breakup mechanisms. However, Sauer et al. \cite{Sauer2014,Sauer2016} and Warncke et al. \cite{Warncke2017} introduced the concept of embedded DNS (eDNS). In this methodology some precursor LES are carried out in order to generate realistic boundary conditions for the gas phase at the DNS inlet. This way, the turbulent fluctuations of velocities are stored in planes that serve as input data for the final target simulation. Warncke et al. \cite{Warncke2019} specifically showed a strong influence of such turbulent inflow condition on the primary breakup process. The method used in the latter eDNS works to solve the gas-liquid interface is the Volume Of Fluid (VOF), in which an indicator function represents the liquid volume contained in a cell \cite{Torregrosa2020,Payri2022}. The advantage of this method is its inherent mass conservation, at the cost of interface smearing. Carmona et al. \cite{Carmona2021} used the incompressible solver NGA combining the VOF method with the Piecewise Linear Interface Calculation (PLIC) technique for interface reconstruction to study the KIT-ITS configuration, properly replicating the breakup mechanism while mantaining a limited CPU consumption.

Level-Set methods have also been used by other authors, but hybridizing them with the VOF method in the so-called CLSVOF \cite{Zandian2017, Agbaglah2017, Mukundan2019a, Mukundan2022} or with the Moment of Fluid method (CLSMOF) \cite{Mukundan2019,Mukundan2022} in order to counter the mass conservation limitations of the former and capture interface topology. 

On the other hand, Lagrangian methods have also been used to model primary breakup in prefilming airblast atomizers. In particular, Smoothed Particle Hydrodynamics (SPH) has been successfully used in 2D \cite{Koch2017, Holz2018} and 3D \cite{Braun2019} planar configurations. Being a meshfree method, it allows a significantly lower computational cost than grid-based simulations. This advantage is reduced when a high number of particles need to be tracked. As a drawback, the interpolation accuracy is influenced by particle arrangement.

The main objective of the present work is to develop a numerical workflow to account for the spatio-temporal variations of the liquid film thickness on a planar prefilmer at the inlet of the atomizer edge DNS, analyzing their effect on primary atomization predictions. To this end, the eDNS concept from Sauer et al. \cite{Sauer2014} is first adopted as a baseline configuration to account for gas inflow turbulence. Section \ref{sec:workflow_case} sketches this numerical workflow explaining how this concept is extended by adding the computation of two-phase precursor LES of the flow on the prefilmer. The particular case of study is also explained in this section. Then, Section \ref{sec:methodology} describes the details for the numerical resolution of each simulation. Additionally, this section details the post-processing methods used to extract information from the ligaments and droplets shaped during breakup. This includes a method to characterize the ligaments not only in the streamwise direction (usual procedure to validate against literature shadowgraphy data), but in the 3D domain. This approach constitutes another main contribution of the investigation to the state of the art. Results are shown in Section \ref{sec:results}, including the validation of the precursor simulations and a discussion on the DNS data from a qualitative and a quantitative standpoint. Finally, the main findings of the manuscript are described in Section \ref{sec:conclusions}.

\section{Proposed simulation workflow and case study} \label{sec:workflow_case}

\begin{figure}[htbp]
	\centering
	\includegraphics[width=1\textwidth]{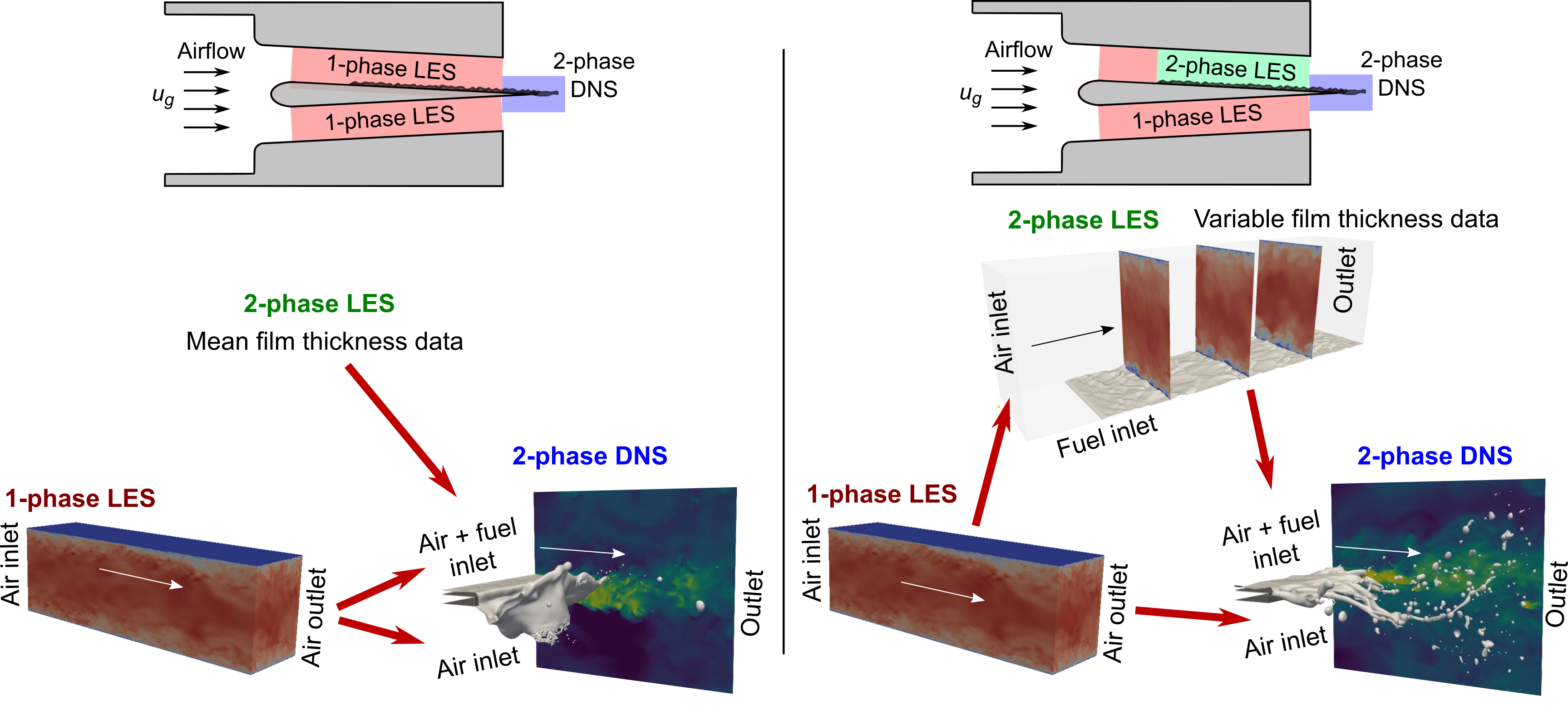}
	\centering
	\caption{Simulation workflow to account for gas inflow turbulence and a constant liquid film thickness at the DNS inlet (left) and proposed workflow to account for the liquid film thickness evolution at the DNS inlet (right).}
	\label{fig:simulationWorkflow}
\end{figure}

The particular planar airblast atomizer to be studied was presented in Figure \ref{fig:KITTestRig}, corresponding to the KIT-ITS test rig \cite{Gepperth2010}. As stated in Section \ref{sec:introduction}, this investigation relies on the eDNS concept by Sauer et al. \cite{Sauer2014} to account for gas inflow turbulence at the atomizing edge DNS as a starting point, with a constant liquid film thickness at the DNS inlet. Figure \ref{fig:simulationWorkflow} (left) depicts the simulation workflow used to replicate this approach, where it can be seen that a precursor single-phase LES is needed reproducing the channel flow above and below the prefilmer surface. Time-varying velocity data from this simulation are mapped to the inlet of the subsequent DNS, restricted to the last part of the prefilmer and a short distance downstream of its edge. This DNS will hereinafter be referred to as \textit{Constant $h_l$ DNS} or \textit{$h_l \neq f(t,z)$ DNS}. Please note that care is taken so that the LES data passed on to both sides of the prefilmer surface of the DNS are not synchronized. Figure \ref{fig:simulationWorkflow} (right), in turn, shows the proposed simulation workflow to extend the method to account for temporal and spatial liquid film waves on the prefilmer that could influence breakup. In this case, the single-phase LES velocity data only feed the lower part of the DNS inlet, being also mapped to an additional two-phase flow precursor LES that simulates the film evolution above the prefilmer. Hence, the liquid boundary layer is developed as it would do through the prefilmer surface. Time-varying velocity and liquid volume fraction results of this simulation are then mapped to the upper part of the DNS inlet obtaining not only a variable airstream velocity but also a variable velocity and height of the liquid film. Consequently, such DNS will hereinafter be referred to as \textit{Time-varying $h_l$ DNS} or \textit{$h_l = f(t,z)$ DNS}. This information is synthesized in Table~\ref{tab:cases} for future reference.

\begin{table}[ht]
\caption{Cases of study corresponding to the strategies used to supply data to the inlet boundary condition of the atomizing edge DNS.}
\centering
\begin{tabular}{ccccc} 
\hline
Case & Description & Air velocity & Liquid velocity & Film thickness \\
\hline
$h_l \neq f(t,z)$ & Constant $h_l$ DNS & Variable & Constant & Constant \\ %\hline
$h_l = f(t,z)$ & Time-varying $h_l$ DNS & Variable & Variable & Variable \\ \hline
\end{tabular}
\label{tab:cases}
\end{table}

Both cases focus on one of the reference geometries of the KIT-ITS test rig \cite{Gepperth2010}, whose parameters are summarized in Table \ref{tab:KITgeometry}. Even though the fuel is injected through an array of discrete holes distributed along the span on the prefilmer upper surface, it results in a totally wetted prefilmer for all the operating conditions experimentally reported in the literature, with an effective prefilming length $L_f$.

\begin{table}[ht]
\caption{Geometrical parameters of the planar prefilmer from the KIT-ITS test rig \cite{Gepperth2010}.}
\centering
\begin{tabular}{llS[table-format=2.2]} 
\hline
Description & Parameter & {Value [mm]} \\
\hline
Length & $L_p$ & 70.9 \\ %\hline
Span & $b$ & 50 \\
Edge thickness & $h_p$ & 0.23 \\
Channel height & $h_c$ & 8.11 \\
Effective film length & $L_f$ & 47.6 \\ \hline
\end{tabular}
\label{tab:KITgeometry}
\end{table}

Both studied cases are investigated for the operating condition detailed in Table \ref{tab:operatingCondition}, corresponding to the test rig operated at atmospheric conditions. The reason for the choice of this condition, representative of altitude relight \cite{Mosbach2010}, is that it has been extensively studied in the literature both experimentally \cite{Gepperth2010, Gepperth2012, Chaussonnet2018, Holz2018} and numerically \cite{Mukundan2019a, Braun2019, Warncke2017, Warncke2019, Carmona2021, Mukundan2019, Mukundan2022}. Hence, a large database is consolidated and readily available to validate the developed numerical methodology.

Air is used as a gas in this operating condition, whereas the liquid is Shellsol D70. This liquid was chosen in the experiments to replicate the properties of kerosene or Jet A-1 at engine-like operating conditions. The relevant non-dimensional groups detailed in Table \ref{tab:operatingCondition} are estimated as follows. The Reynolds numbers ${Re}$ for each phase are calculated according to Eq. (\ref{eq:Re}):
\begin{equation}
    {Re}_g = \cfrac{\rho_g \, \bar{u}_g \, h_c / 2}{\mu_g} \quad \text{and} \quad {Re}_l = \cfrac{\rho_l \, \dot{V}/b}{\mu_l}
\label{eq:Re}
\end{equation}

where it can be seen that the half-channel width is chosen as the reference length for ${Re}_g$ and the average liquid film thickness is used for ${Re}_l$ from the volumetric flow rate or film loading $\dot{V}/b$.

As far as the Weber numbers are concerned, their definitions are in Eq. (\ref{eq:We}):
\begin{equation}
    {We}_l = \cfrac{\rho_g \, \left( \bar{u}_g - \bar{u}_l \right)^2 \, h_l}{\sigma} \qquad {We}_{hp} = \cfrac{\rho_g \, \bar{u}_g^2 \, h_p}{\sigma} \quad \text{and} \quad {We}_{\delta} = \cfrac{\rho_g \, \bar{u}_g^2 \, \delta_{TE}}{\sigma}
\label{eq:We}
\end{equation}

where the main differences among definitions come from the characteristic dimension considered. For ${We}_l$, $h_l$ is estimated from the volumetric flow rate and later verified in the computed 2-phase LES. The interest of ${We}_{hp}$ resides in the fact that Chaussonnet et al. found that the SMD in part scales with $1 / \sqrt{We_{hp}}$ \cite{Chaussonnet2016}. On the other hand, $We_{\delta}$ is based on the boundary layer thickness at the prefilmer trailing edge $\delta_{TE}$, which is in turn estimated according to Eq. (\ref{eq:deltaTE}) by analogy to a turbulent boundary layer on a flat plate \cite{White2006}:
\begin{equation}
    \delta_{TE} = 0.16 \, \cfrac{L_p}{Re_{Lp}^{1/7}}
\label{eq:deltaTE}
\end{equation}

where $Re_{Lp}$ is a specific gaseous $Re$ based on the prefilmer length as the relevant dimension, which takes a value of 158,230 for the operating condition tested.

\begin{table}[ht]
\caption{Functional parameters of the studied operating condition.}
\centering
\begin{tabular}{llS[table-format=5.4]l} 
\hline
Description & Parameter & {Value} & Units \\
\hline
Temperature & $T$ & 298 & K \\ %\hline
Pressure & $p$ & 1 & atm \\
\hline
Gas bulk velocity & $\bar{u}_g$ & 50 & m/s \\ %\hline
Gas density & $\rho_g$ & 1.21 & kg/m\textsuperscript{3} \\
Gas dynamic viscosity & $\mu_g$ & \SI{1.82e-5}{} & Pa·s \\
\hline
Liquid normalized volumetric flow rate & $\dot{V}/b$ & 50 & mm\textsuperscript{2}/s \\ %\hline
Liquid density & $\rho_l$ & 770 & kg/m\textsuperscript{3} \\
Liquid dynamic viscosity & $\mu_l$ & \SI{1.56e-3}{} & Pa·s \\
Liquid surface tension & $\sigma$ & 0.0275 & kg/s\textsuperscript{2} \\
\hline
Gas Reynolds number & ${Re}_g$ & 13480 & - \\ %\hline
Liquid Reynolds number & ${Re}_l$ & 24.63 & - \\ %\hline
Weber number (traditional) & ${We}_l$ & 8.6 & - \\ %\hline
Weber number (based on atomizing edge) & ${We}_h$ & 25.3 & - \\ %\hline
Weber number (based on boundary layer) & ${We}_{\delta}$ & 151.5 & - \\ %\hline
Ohnesorge number & $Oh$ & 0.0223 & - \\ %\hline
Momentum flux ratio & $M$ & 15.71 & - \\ \hline
\end{tabular}
\label{tab:operatingCondition}
\end{table}

\section{Computational methodology} \label{sec:methodology}

The present section covers the details of the numerical simulations needed to complete the computational workflows depicted in Figure \ref{fig:simulationWorkflow} and Table \ref{tab:cases}. Hence, it presents the governing equations, meshing strategy and computational setup for each of the precursor LES needed to generate reliable data for the embedded DNS of the atomizing edge and for the DNS itself. The methodology used to post-process both the ligament arrangement and the droplet cloud predicted by the DNS are also explained in this section.

\subsection{Precursor single-phase flow LES} \label{subsec:LESmono}

\subsubsection{Governing equations and related submodels}
To solve the transient incompressible and turbulent channel flow along the planar prefilmer duct, the Pressure-Implicit with Splitting of Operators (PISO) algorithm \cite{Issa1986} is used through the \textit{pisoFoam} solver in the OpenFOAM open source CFD toolbox \cite{OpenFOAM}. It solves Eq. (\ref{eq:1phaseLES_continuity}) and Eq. (\ref{eq:1phaseLES_momentum}) as the filtered governing equations for continuity and momentum conservation, respectively:
\begin{equation}
\nabla \cdot \bar{\vb{u}} = 0
\label{eq:1phaseLES_continuity}
\end{equation}
\begin{equation}
\frac{\partial \bar{\vb{u}}}{\partial t} + \nabla \cdot \left( \bar{\vb{u}} \bar{\vb{u}} \right) = - \frac{1}{\rho} \nabla \bar{P} + \nabla \cdot \left[ \left( \nu + \nu_{SGS} \right) \nabla \bar{\vb{u}} \right] \label{eq:1phaseLES_momentum}
\end{equation}

Closure on the filtered governing equations is achieved through the subgrid scale viscosity $\nu_{SGS}$, which is modelled by the Wall Adapting Local Eddy-viscosity (WALE) sub-grid scale (SGS) model from Nicoud and Ducros \cite{Nicoud1999}. 

% WALE por defecto: Ce=1.048 | Ck=0.094 | Cw=0.325

\subsubsection{Computational domain and mesh}
\begin{figure}[htbp]
	\centering
	\includegraphics[width=1\textwidth]{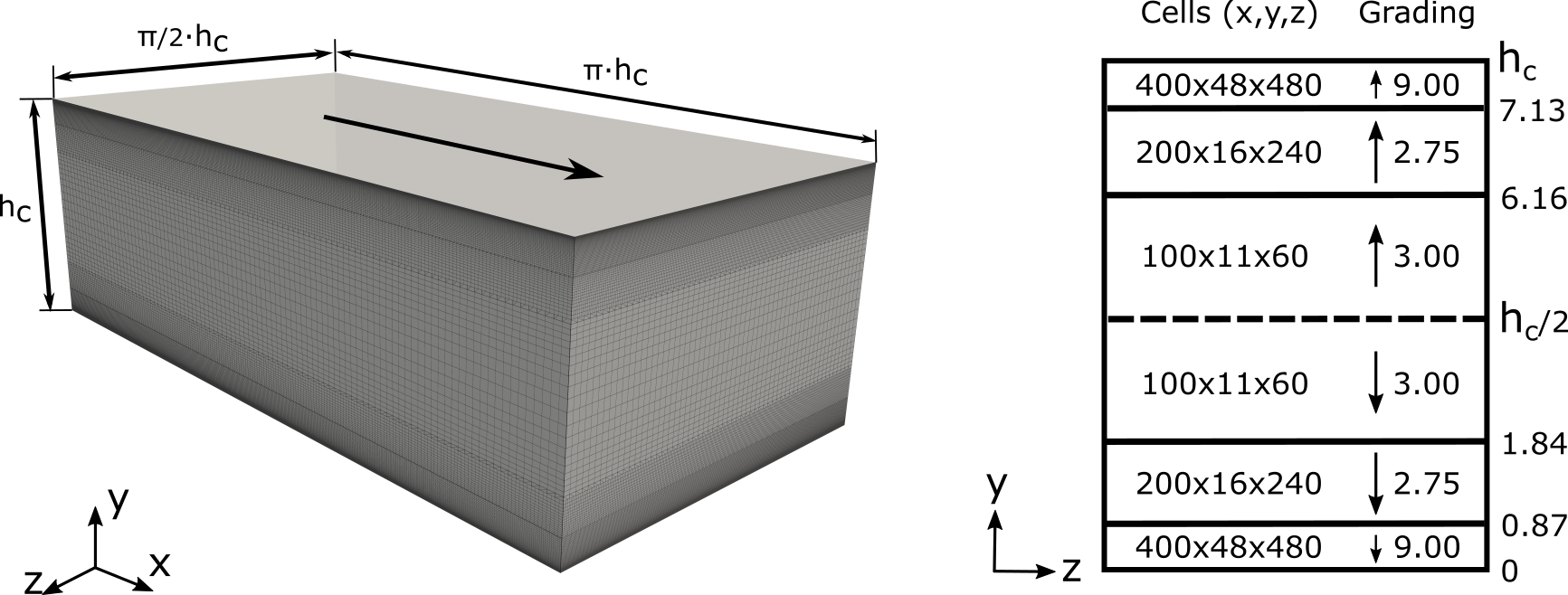}
	\centering
	\caption{Geometry and mesh sketch for the single-phase turbulent channel flow LES (left), detail of the number of cells and cell grading in each of the mesh zones (right).}
	\label{fig:1phaseLES_domain_mesh}
\end{figure}

%% Domain
Figure \ref{fig:1phaseLES_domain_mesh} shows the geometrical dimensions of the computational domain and the mesh for the turbulent channel. The domain height in the wall-normal $y$ direction replicates the full prefilmer channel height from the KIT-ITS test rig, $h_c$. The width in the spanwise $z$ direction is chosen as $\pi /2 \cdot h_c$, according to Moser et al. \cite{Moser1999}. The domain length in the streamwise $x$ direction is $\pi \cdot h_c$, even though a periodic boundary condition is set in the streamwise edges in order to virtually mimic an infinitely long channel.

%% Mesh
A zonal mesh strategy with hexahedral cells is adopted as in the work by Sauer et al. \cite{Sauer2016}. Figure \ref{fig:1phaseLES_domain_mesh} (left) shows the geometry of the domain and the direction of the flow. The grid spacing is uniform in the streamwise and spanwise directions for every mesh zone, but a symmetric cell-size grading is applied in the wall-normal $y$ direction to refine the domain from the channel core to the walls. Figure \ref{fig:1phaseLES_domain_mesh} (right) presents the number of cells and a grading coefficient for all the 3 zones. This grading coefficient $g_c$ is the proportion between the sizes of the first and last cell in that zone, following the arrow direction: $g_c = \frac{cell_0}{cell_N}$. The intermediate cell size increases or decreases linearly. A $y^{+} < 1$ condition is fulfilled at the walls, preventing the use of wall functions. This strategy yields a total of 20.1 M cells.%, with sizes between _ and _..

\subsubsection{Boundary conditions, discretization schemes and solution control}
%% Boundaries and samples
As far as the boundary conditions are concerned, gas velocity is initially prescribed at the inlet (X- boundary) by a Dirichlet boundary condition with $\bar{u}_g=\SI{50}{\meter/\second}$ according to Table \ref{tab:operatingCondition} and null values for $v$ and $w$; whereas a Neumann boundary condition is used at the outlet (X+ boundary). A no-slip condition is used at the bottom and top walls (Y- and Y+ boundaries) and a periodic condition is used spanwise (Z- and Z+ boundaries). The \textit{boxTurb} tool from OpenFOAM is used to initially trigger turbulence in the domain. After a transient, both the inlet and outlet boundaries (X- and X+) are also converted into periodic boundary conditions. This configuration mimics an infinitely long turbulent channel where turbulence can develop to simulate the prefilmer channel and may be passed on to the subsequent simulations according to Figure \ref{fig:simulationWorkflow} workflow. To this end, velocity data are sampled at the central YZ plane of the domain every \SI{1}{\micro\second}. It must be noted that the resultant turbulent channel flow simulation corresponds to ${Re}_{\tau} \approx 685$.

%% Schemes
All chosen discretization schemes are 2\textsuperscript{nd} order in space and time (Gauss linear central differencing and backward, respectively). A Geometric Agglomerated Algebraic Multi-Grid (GAMG) is used as the linear solver. The time step is let to vary during simulation runtime, being set by restricting the maximum CFL number to 0.2. This results in time steps in the order of \SI{1e-7}{\second}.

\subsection{Precursor two-phase flow LES} \label{subsec:LESbi}

\subsubsection{Governing equations and related submodels}
%%%%% Phisics
In order to solve the transient two-phase flow with incompressible, isothermal and immiscible fluids, the \textit{interFoam} solver within OpenFOAM is used. It applies the Volume of Fluid (VOF) interface capturing approach, which introduces the cell liquid volume fraction $\alpha = V_l / V_{cell}$ as a variable and advects it according to Eq.~\ref{eq:2phaseLES_alphaAdvection}.
\begin{equation}
\frac{\partial \bar{\alpha}}{\partial t} + \bar{\vb{u}} \cdot \nabla \bar{\alpha} + \bar{\vb{u_c}} \cdot \nabla \left[ \bar{\alpha} \left( 1 - \bar{\alpha} \right) \right]  = 0 
\label{eq:2phaseLES_alphaAdvection}
\end{equation}

The last term in Eq. (\ref{eq:2phaseLES_alphaAdvection}) is an artificial compression term that only acts in the vicinity of the interface, creating a flux that counters numerical diffusion keeping a sharp interface \cite{Wardle2013}. The velocity vector $\bm{u_c}$ is defined according to Eq. (\ref{eq:2phaseLES_compressionVelocity}):
\begin{equation}
 \vb{u_{c}} = C_{\alpha} \, \lvert \vb{u} \rvert \, \vb{n}
\label{eq:2phaseLES_compressionVelocity}
\end{equation}

where $C_{\alpha}$ is a binary coefficient that switches interface sharpening on (with a value of 1) or off (with a value of 0), whereas $\vb{n}$ is the interface unit normal vector, used to define the direction of the applied compression velocity and approximated by Equation \ref{eq:2phaseLES_normalVector}:
\begin{equation}
 \vb{n} = \cfrac{\nabla \alpha}{\lvert \nabla \alpha \rvert}
\label{eq:2phaseLES_normalVector}
\end{equation}

While this interface compression method is in general not as physically accurate as interface reconstruction methods (such as PLIC), it is mass conservative \cite{Wardle2013}. This is the reason behind its choice for this precursor simulation.

Additionally, surface tension is modelled through the Continuum Surface Force (CSF) model \cite{Brackbill1992}. A surface tension force $\vb{f_{\sigma}}$ is then included (correspondingly divided by $\rho$) as a source term in the momentum equation (\ref{eq:1phaseLES_momentum}), depending on the surface tension $\sigma$ and the interface curvature $\kappa$:
\begin{equation}
 \vb{f_{\sigma}} = \sigma \, \kappa \, \delta_S \, \vb{n}
\label{eq:2phaseLES_momentumSource}
\end{equation}

where $\delta_S$ is a Dirac delta function that concentrates the effects of the surface tension on the liquid interface only and the curvature is approximated through Eq. \ref{eq:2phaseLES_curvature}:
\begin{equation}
 \kappa = - \nabla \cdot \vb{n}
\label{eq:2phaseLES_curvature}
\end{equation}

The interface unit normal vector in Eq. (\ref{eq:2phaseLES_momentumSource}) and (\ref{eq:2phaseLES_curvature}) is again approximated through Eq. (\ref{eq:2phaseLES_normalVector}).

The density $\rho$ and viscosity $\mu$ at a cell used in the continuity and momentum conservation equations are linear expressions of the fluid properties weighted by $\alpha$.
\begin{equation}
\begin{split}
\rho = \alpha\rho_l + (1-\alpha)\rho_g \\
\mu = \alpha\mu_l + (1-\alpha)\mu_g
\end{split}
\label{eq:2phaseLES_densvisc}
\end{equation}

Last, the WALE SGS model is again used to model sub-grid turbulence.

%%%%% Computational setup

\subsubsection{Computational domain and mesh}
\begin{figure}[htbp]
	\centering
	\includegraphics[width=1\textwidth]{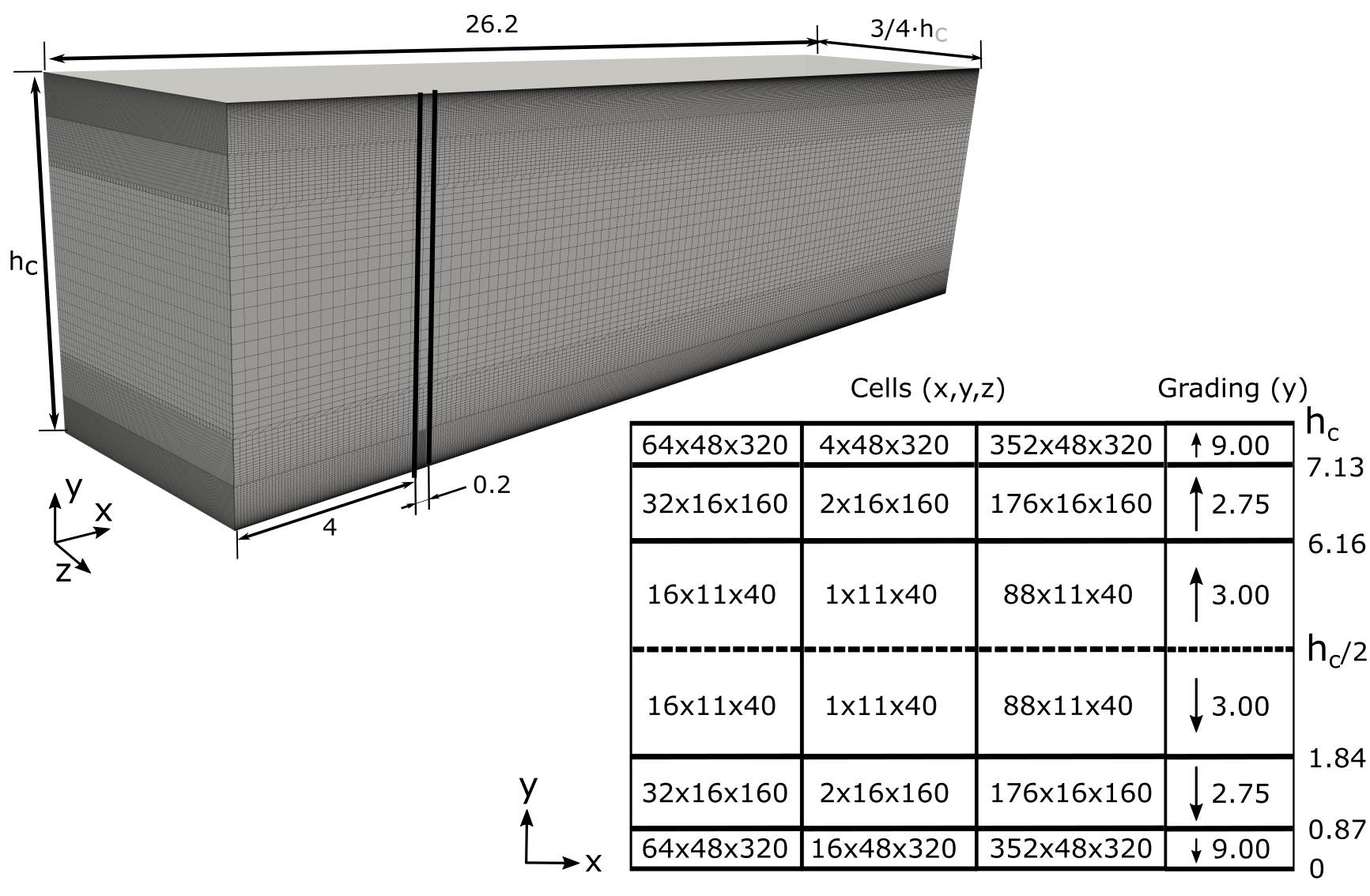}
	\centering
	\caption{Geometry and mesh sketch for the two-phase turbulent channel flow LES.}
	\label{fig:2phaseLES_domain_mesh}
\end{figure}

%% Domain
Figure \ref{fig:2phaseLES_domain_mesh} shows the geometry considered for the computational domain of the two-phase channel flow, together with the mesh used. The domain dimensions keep the full channel height $h_c$ in the wall-normal $y$ direction. In the streamwise $x$ direction, there is enough space from the fuel inlet (included at the bottom of the domain) to the domain outlet for the film to develop half of the prefilmer effective film length $L_f$. Including some space to separate the fuel inlet from the air inlet, this yields a total length of \SI{26.2}{\milli\meter}. Since the mesh in this case is finer than the one for the single-phase flow precursor LES, the width in the spanwise $z$ direction is chosen as $3/4 \cdot h_c$ (nearly half of the single-phase flow precursor LES) in order to keep a limited amount of computational resources.

%% Mesh
A zonal mesh strategy is utilized again. A cell-size grading is prescribed in the wall-normal $y$ direction, similar to the one used in the single-phase flow precursor LES. In this case, however, an additional refined zone is added at the lower part of the domain in order to better represent the gas-liquid interface. This yields a 4-step sequence from the wall-normal domain center to the bottom, keeping the condition of $y^{+} < 0.9$ at the top and bottom walls, again preventing the use of wall functions. A local refinement is added at the streamwise location where the fuel inlet is placed. The grid spacing is uniform in the spanwise direction, but a slight grading is applied streamwise towards the outlet. This strategy yields a total cell count of 21.4 M cells.

\subsubsection{Boundary conditions, discretization schemes and solution control}
%% Boundaries and samples
Air velocity is provided at the gas inlet (X- boundary) as a Dirichlet boundary condition with time-varying data mapped from the single-phase LES simulation, linearly interpolating among sampled planes. The fuel inlet at the Y- boundary occupies part of the streamwise direction and is uniformly distributed spanwise, so as to replicate the uniform fuel injection achieved upstream of the prefilmer edge at all operating conditions in the KIT-ITS test rig. A Dirichlet boundary condition is also used, liquid velocity being imposed according to the inlet area in order to prescribe the flow rate from Table \ref{tab:operatingCondition}. A Neumann boundary condition is used at the domain outlet (X+ boundary). A no-slip condition is used at the bottom (except for the fuel inlet region) and top walls (Y- and Y+ boundaries) and a periodic condition is again used spanwise (Z- and Z+ boundaries).

This configuration mimics the upper part of the KIT-ITS planar prefilmer so both the liquid film waves and the air-fuel boundary layer may be passed on to the subsequent atomizing edge DNS according to Figure \ref{fig:simulationWorkflow} (right) workflow. To this end, velocity and liquid volume fraction data are sampled at a YZ plane close to the domain outlet every \SI{2}{\micro\second}.

%% Schemes
The pressure-velocity coupling is again solved using the PISO algorithm. The Multidimensional Universal Limiter for Explicit Solution (MULES) method within \textit{interFoam} is used to solve for the liquid volume fraction advection given by Eq. (\ref{eq:2phaseLES_alphaAdvection}). This method uses a Flux-Corrected Transport (FCT) algorithm \cite{Zalesak1979} and ensures the boundedness of the solution.

As it happened for the single-phase precursor LES, the chosen discretization schemes are 2\textsuperscript{nd} order in time and space. The Crank-Nicolson method is used for the temporal discretization of $\alpha$ in Eq. (\ref{eq:2phaseLES_alphaAdvection}). The Geometric Agglomerated Algebraic Multi-Grid (GAMG) is again used as the linear solver. A fixed time step of \SI{5e-8}{\second} is used in this case, ensuring the maximum CFL number never surpasses 0.25.

\subsection{Atomizing edge two-phase flow DNS} \label{subsec:DNS}

\subsubsection{Governing equations and related submodels}
%%%%% Phisics
To solve the incompressible, isothermal and immiscible two-phase flow at the atomizing edge with a Direct Numerical Simulation (DNS) approach, the PARIS (PArallel, Robust, Interface Simulator) code \cite{Aniszewski2021} is used. It applies the Navier–Stokes equations in the one-fluid formulation of two-phase flow, including also the Continuous Surface Force (CSF) method to compute surface tension forces:
\begin{equation}
 \rho \, \left[ \frac{\partial \vb{u}}{\partial t} + \nabla \cdot \left( \vb{u} \vb{u} \right) \right] = - \nabla p + \mu \nabla^2 \vb{u} + \sigma \, \kappa \, \delta_S \, \vb{n}
\label{eq:2phaseDNS_NavierStokes}
\end{equation}

The VOF method is used to capture the interface according to Eq. (\ref{eq:2phaseDNS_alphaAdvection}), computing density and viscosity as in Eq. (\ref{eq:2phaseLES_densvisc}).
\begin{equation}
\frac{\partial \alpha}{\partial t} + \vb{u} \cdot \nabla \alpha = 0 
\label{eq:2phaseDNS_alphaAdvection}
\end{equation}

Please note the differences in VOF advection with the two-phase precursor LES. In this case, a modified Piecewise Linear Interface Capturing (PLIC) approach called CIAM \cite{Li95} is used to reconstruct the interface and to compute its unit normal vector $\vb{n}$ \cite{Scardovelli2003} for the CSF. The curvature $\kappa$ is estimated from the Height Function (HF) method proposed by Popinet \cite{Popinet2009} and whose full implementation details in PARIS are described by Aniszewski et al. \cite{Aniszewski2021}. For these reasons, Eq. (\ref{eq:2phaseDNS_alphaAdvection}) leaves aside the artificial interface compression term introduced in Eq. (\ref{eq:2phaseLES_alphaAdvection}).

%%%%% Computational setup
\subsubsection{Computational domain and mesh}

\begin{figure}[htbp]
	\centering
	\includegraphics[width=0.45\textwidth]{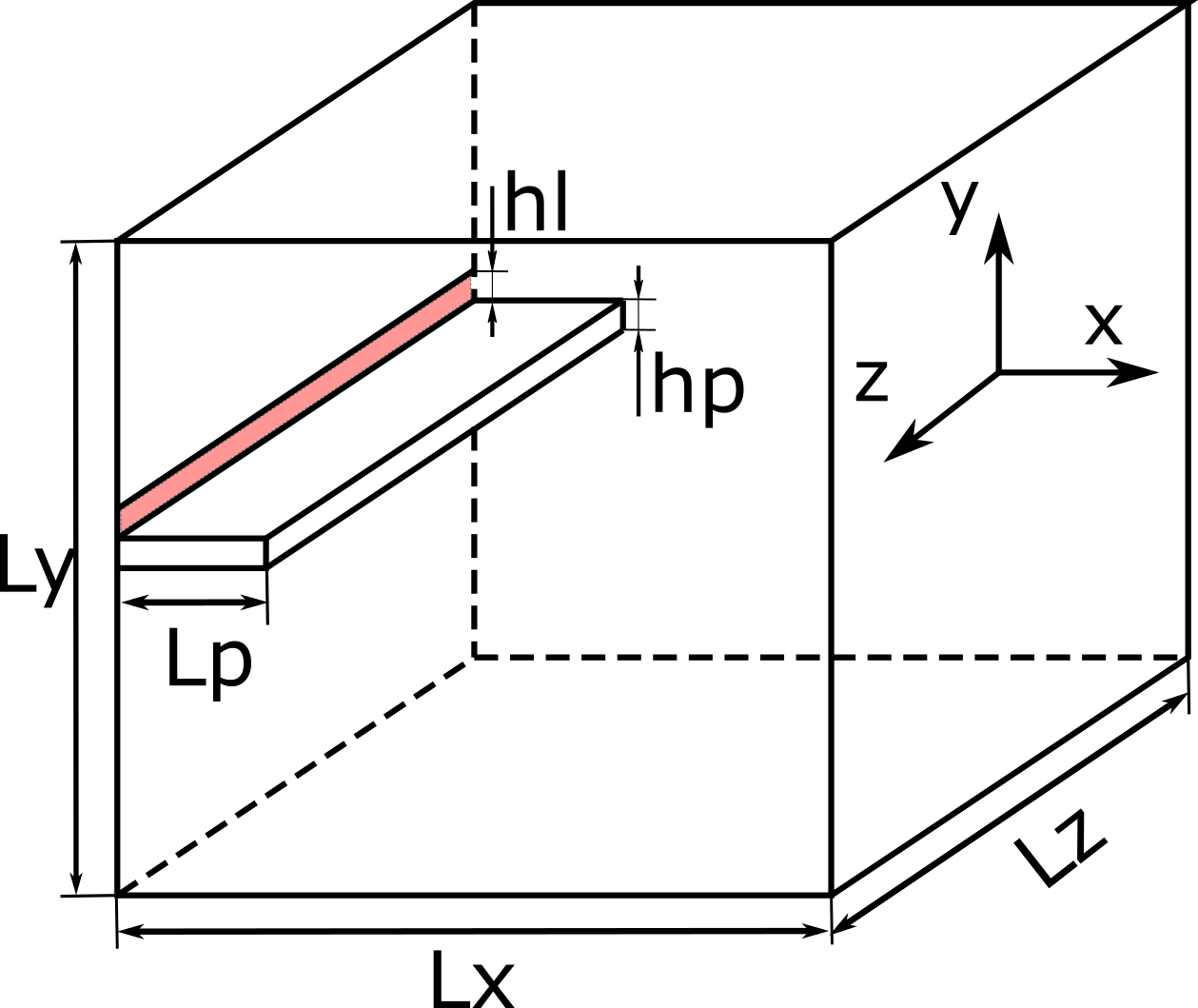}
%	\centering
	\caption{Atomizing edge DNS computational domain. Flow direction is $X^+$ (from left to right).}
	\label{fig:case_DNSdomain}
\end{figure}

The geometry of the atomizing edge DNS domain is showed in Figure \ref{fig:case_DNSdomain}. The illustrated domain dimensions are compiled in Table \ref{tab:DNSgeometry}. These dimensions are very similar to the ones used in comparable computations in the literature \cite{Mukundan2019a,Sauer2014,Warncke2017,Mukundan2019,Mukundan2022}. Additionally, the prefilmer edge thickness $h_p$ was shown in Table \ref{tab:KITgeometry}. As per the liquid film thickness, the value $h_l = \SI{0.08}{\milli\meter}$ is used in the $h_l \neq f(t,z)$ case (corresponding to the rounded value obtained in the precursor two-phase LES), whereas it fluctuates in the $h_l = f(t,z)$ case as stated in Table \ref{tab:cases}, this being the object of investigation.

\begin{table}[ht]
\caption{Main dimensions of the atomizing edge DNS computational domain.}
\centering
\begin{tabular}{llS[table-format=1.2]} 
\hline
Description & Parameter & {Value [mm]} \\
\hline
Domain length & $L_x$ & 6.08 \\ %\hline
Domain height & $L_y$ & 6.08 \\
Domain span & $L_z$ & 4.00 \\
Prefilming length & $L_p$ & 1.00 \\ \hline
\end{tabular}
\label{tab:DNSgeometry}
\end{table}

%%%%% Mesh
A staggered Cartesian grid is used in this case, with pressure collocated at the cell centres and velocities located at the face centres. The calculated $Re$ yields a Kolmogorov length scale $l_{\eta} = \SI{12}{\micro\meter}$, implying a minimum grid spacing of $\Delta x_{min} = \SI{25}{\micro\meter}$ according to Pope's criterion \cite{Pope2000}. A cell size $\Delta x = \Delta y = \Delta z = \SI{10}{\micro\meter}$ is chosen for both cases, resulting in 147.87 million cells. This resolution matches the one by Warncke et al. \cite{Warncke2017}, who already justified that Sauer et al. \cite{Sauer2016} had performed a detailed discussion about the droplet representation with different grid sizes ($\SI{20}{\micro\meter}$ and $\SI{10}{\micro\meter}$). They found that a resolution of the entire liquid mass is hardly achievable, but that the $\SI{10}{\micro\meter}$ cell size allowed a resolution of $80\%$ of the liquid mass.
 
\subsubsection{Boundary conditions, discretization schemes and solution control}

A Dirichlet boundary condition is provided at the inlet (X- boundary), with time-varying velocity data mapped from the precursor LES in the $h_l \neq f(t,z)$ case, and both time-varying velocity and liquid volume fraction $\alpha$ in the $h_l = f(t,z)$ case. Linear interpolations among temporally sampled planes are used in both cases. A special convective and time-varying boundary is used in the outlet (X+ boundary) to reduce reflections \cite{Kajishima2018}, as expressed in Eq. (\ref{eq:2phaseDNS_convectiveOutletBC}), where $u_m$ is computed at each time step as the average streamwise velocity at the inlet.
\begin{equation}
 \frac{\partial u_i}{\partial t} + u_m \frac{\partial u_i}{\partial x} = 0
\label{eq:2phaseDNS_convectiveOutletBC}
\end{equation}

For the bottom and top bounds (Y- and Y+ boundaries) a slip boundary condition (i.e. moving wall) is used with a streamwise velocity equal to the mean gas velocity. This condition allowed containing major flow vortices, with the chosen $L_y$ dimension ensuring no influence on the region of interest for primary atomization. A periodic condition is used spanwise (Z- and Z+ boundaries).

Last, a no-slip boundary condition is used for the immersed boundary (prefilmer solid wall). A static contact angle among the prefilmer solid wall and the liquid has been introduced as a VOF condition to account for wettability. As formulated by Lacis et al. \cite{Lacis2022} following Legendre \& Maglio's work \cite{Legendre2015}, this boundary condition is mathematically set by defining the interface normal unit vector at the boundary ($\vb{n}$) through Eq. (\ref{eq:2phaseDNS_contactAngle}) wherever the interface is attached to the prefilmer wall:
\begin{equation}
 \vb{n} \cdot \vb{n_{wall}} = \cos{\left( \theta \right)}
\label{eq:2phaseDNS_contactAngle}
\end{equation}

where $\vb{n_{wall}}$ is the wall-normal unit vector. A contact angle $\theta = 60^{\circ}$ has been set between the liquid and the prefilmer according to Braun et al. \cite{Braun2019}.

%%%%% PARIS methods
The spatial and time discretization schemes are 2\textsuperscript{nd} order, using a multigrid solver with a V-cycle to solve the Poisson equation. In the case of temporal discretization, a predictor-corrector method is used, with an explicit projection step for the pressure. A complete description of the methods is given elsewhere \cite{Aniszewski2021,tryggvason2011direct}. A variable time step is used (limited to a maximum of \SI{2e-8}{\second}), ensuring the maximum CFL number never surpasses 0.2. Computations are run with 4096 CPUs for a total simulated time of about $\SI{9}{\milli\second}$, so that 3 main breakup events are accounted for in both cases (a first event is discarded, as it corresponds to a simulation transient). Simulations were run with 4096 CPUs, with 0.21M CPUh needed per each millisecond of simulated physical time.

\subsubsection{Post-processing} \label{sec:pospro}

%%%%% Postprocessing
Different post-processing tools have been implemented and applied to the atomizing edge DNS data in order to generate and analyze quantitative results.

\begin{figure}[htbp]
	\centering
	\includegraphics[width=0.8\textwidth]{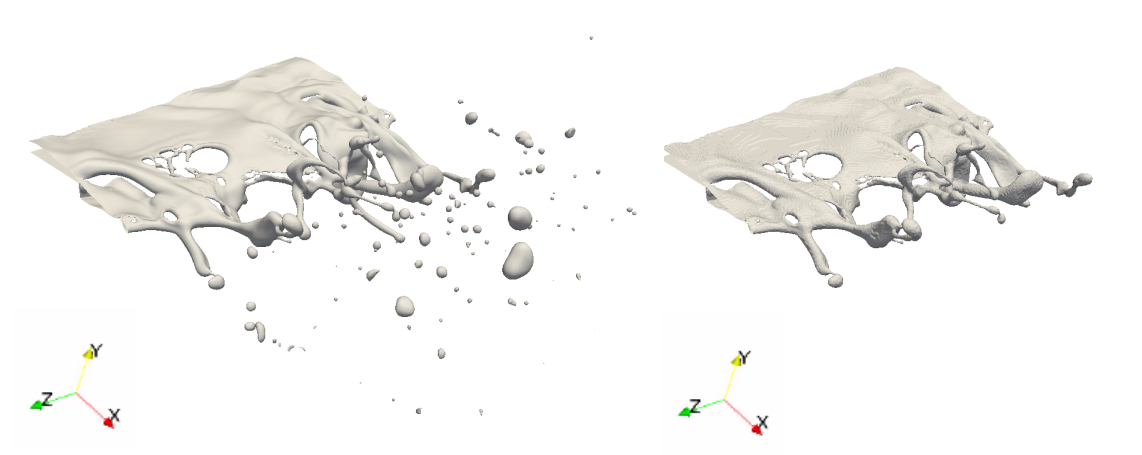}
%	\centering
	\caption{All liquid structures identified in the domain through an iso-surface of $\alpha = 0.5$ (left), liquid core extracted through the connectivity algorithm (right). Images correspond to a given instant of the $h_l = f(t,z)$ case.}
	\label{fig:pospro_separate}
\end{figure}

The first step to post-process the DNS data is to identify each individual continuous liquid structure. This is done scanning the whole domain looking for free surfaces according to an $\alpha$ threshold. Figure \ref{fig:pospro_separate} (left) shows a visualization of the instantaneous liquid-gas interface through an iso-contour of $\alpha = 0.5$ (value used as a threshold in the present investigation). After a binarization using that threshold, the connectivity algorithm from The Visualization Toolkit (VTK) \cite{VTK2018} is applied to the domain for all time steps in order to identify every isolated liquid structure, obtaining their position and size. Once this is achieved, the intact core is identified as the first structure (including the film and any ligament directly attached to it, as shown in Figure \ref{fig:pospro_separate} -right) and separated from the droplet cloud and any detached ligaments. This way, the intact core and ligaments can be independently processed on the one hand, whereas the droplet characteristics can be analyzed on the other hand.

\begin{itemize}
    \item Intact core and ligament analysis
\end{itemize}
%%% LIGAMENTS
As far as the ligaments are concerned, it must be noted that experimental results available in the literature provide 2D shadowgraphy images \cite{Gepperth2012, Chaussonnet2020, Warncke2017}. Hence, as a first approach, the liquid data from the DNS is projected on the XZ plane in order to obtain comparable images for validation purposes.

\begin{figure}[htbp]
	\centering
	\includegraphics[width=1\textwidth]{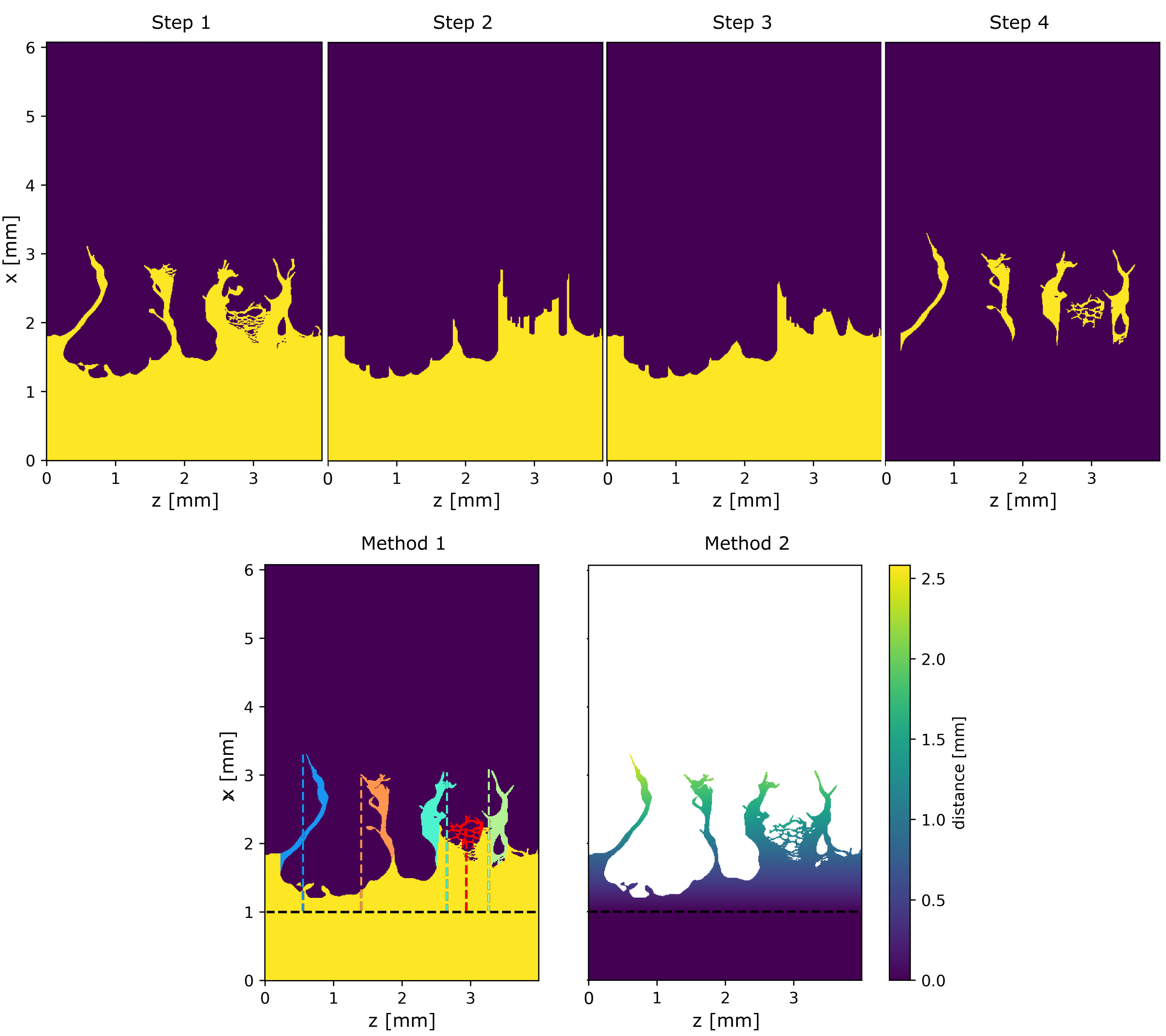}
%	\centering
	\caption{Steps for 2D ligament analysis (top): liquid core and ligaments projected over the XZ plane and binarized (Step 1), mask construction (Steps 2 and 3) and isolated ligaments obtained (Step 4). Methods for ligament length measurement (bottom): Axial distance method (Method 1) and Fast Marching Method (Method 2). Images correspond to a given instant of the $h_l \neq f(t,z)$ case.}
	\label{fig:pospro_ligs_detection}
\end{figure}

The liquid projection is obtained by aggregating the values of the $\alpha$ variable along the Y axis at each XZ plane location. The resulting data is binarized with a chosen threshold ($\alpha = 1$), yielding images such as the one labelled as Step 1 in Figure \ref{fig:pospro_ligs_detection}. Then, the isolated ligaments must be extracted from the intact core or liquid film. The method proposed to this end is as follows. For each Z coordinate, a function expressing the number of liquid cells before the first gas cell along the X axis is computed (Figure \ref{fig:pospro_ligs_detection} Step 2). Next, big slopes are detected with the derivative of this function and erased with a cutting and smoothing algorithm, obtaining the main liquid core except for the elongated structures (Figure \ref{fig:pospro_ligs_detection} Step 3). This structure is then used as a cutting mask, substracting it from the projected domain from Step 1 and obtaining the ligaments as isolated structures, noise being removed by deleting structures with very small areas (Figure \ref{fig:pospro_ligs_detection} Step 4). Finally, each ligament is labeled using the connectivity algorithm from OpenCV \cite{openCV}. 

Once a ligament is characterized, the cell with the maximum X coordinate is considered to be the tip of that ligament. With that information, there are two methods for measuring string length $L_{str2D,i}$ as shown in figure \ref{fig:pospro_ligs_detection} (bottom). Method 1 (left) considers the ligament length as the axial distance among the prefilmer edge and the ligament tip, following a straight line in the X direction. This is the simplest method and is also used in the literature shadowgraphy experiments and the simulations by Warncke et al. \cite{Warncke2017}. Hence, it has been used to validate the computations of the present work. Method 2 (right), proposed in this study, uses the \textit{distance} function of the \textit{scikit-fmm} package in Python, which is an implementation of the Fast Marching Method by Sethian et al. \cite{Sethian1996} to solve an Eikonal equation. Defining the prefilmer edge as a propagating surface, and setting the same constant velocity for every node in the domain, it returns a distance matrix that is equivalent to the continuous shortest path along ligament connectivity points between the prefilmer edge and each ligament cell.

\begin{figure}[htbp]
	\centering
	\includegraphics[width=0.3\textwidth]{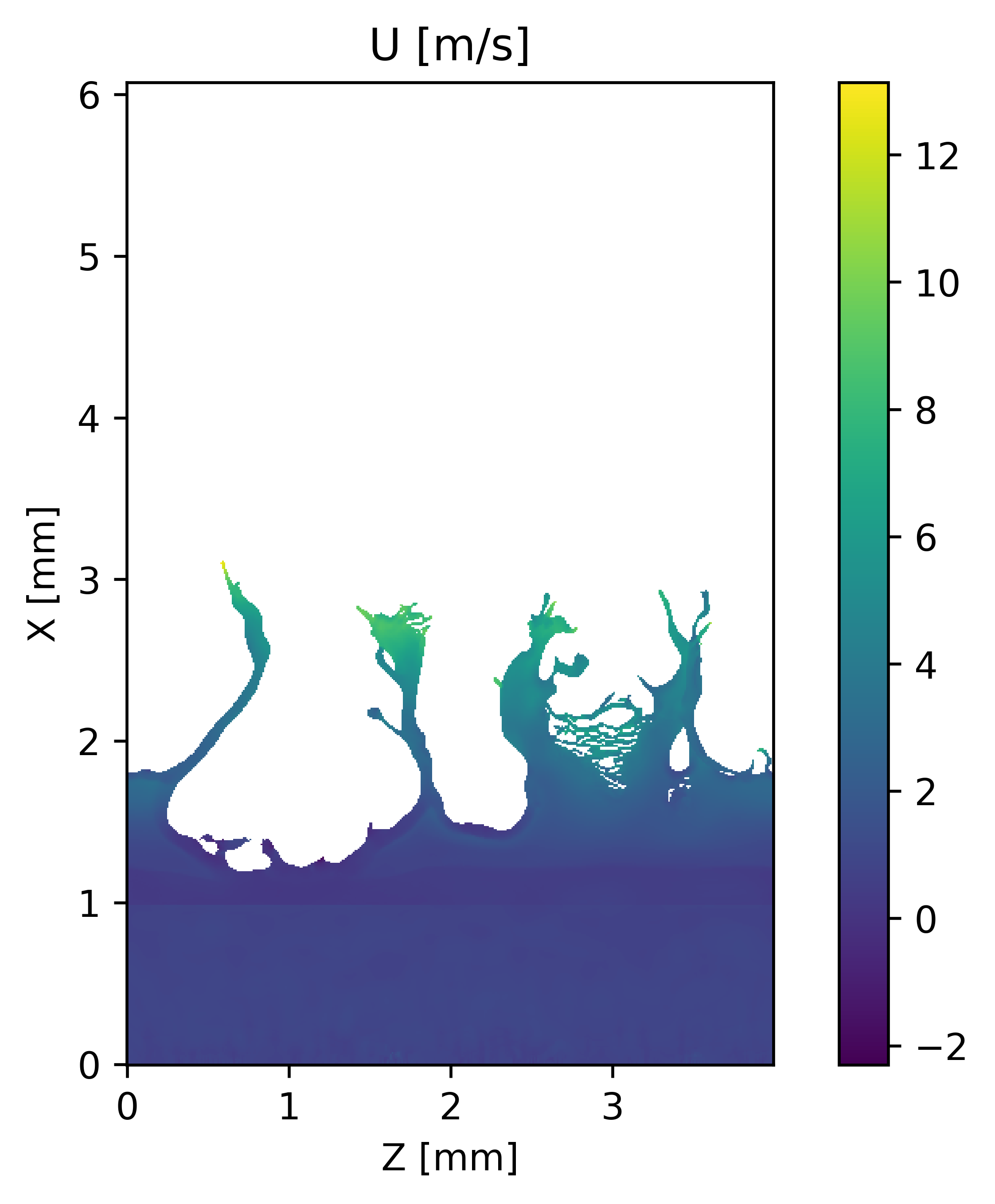}
	\includegraphics[width=0.3\textwidth]{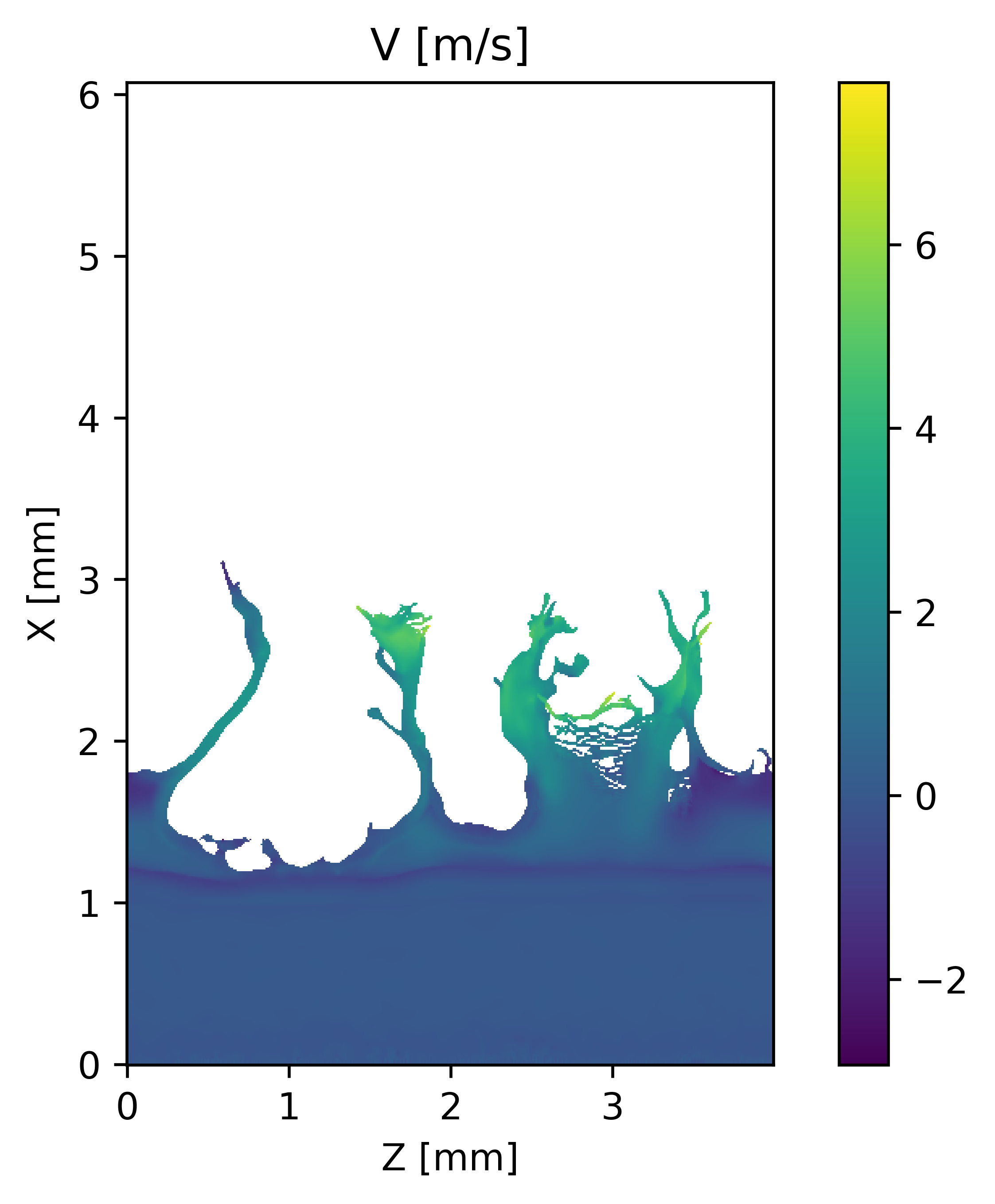}
	\includegraphics[width=0.3\textwidth]{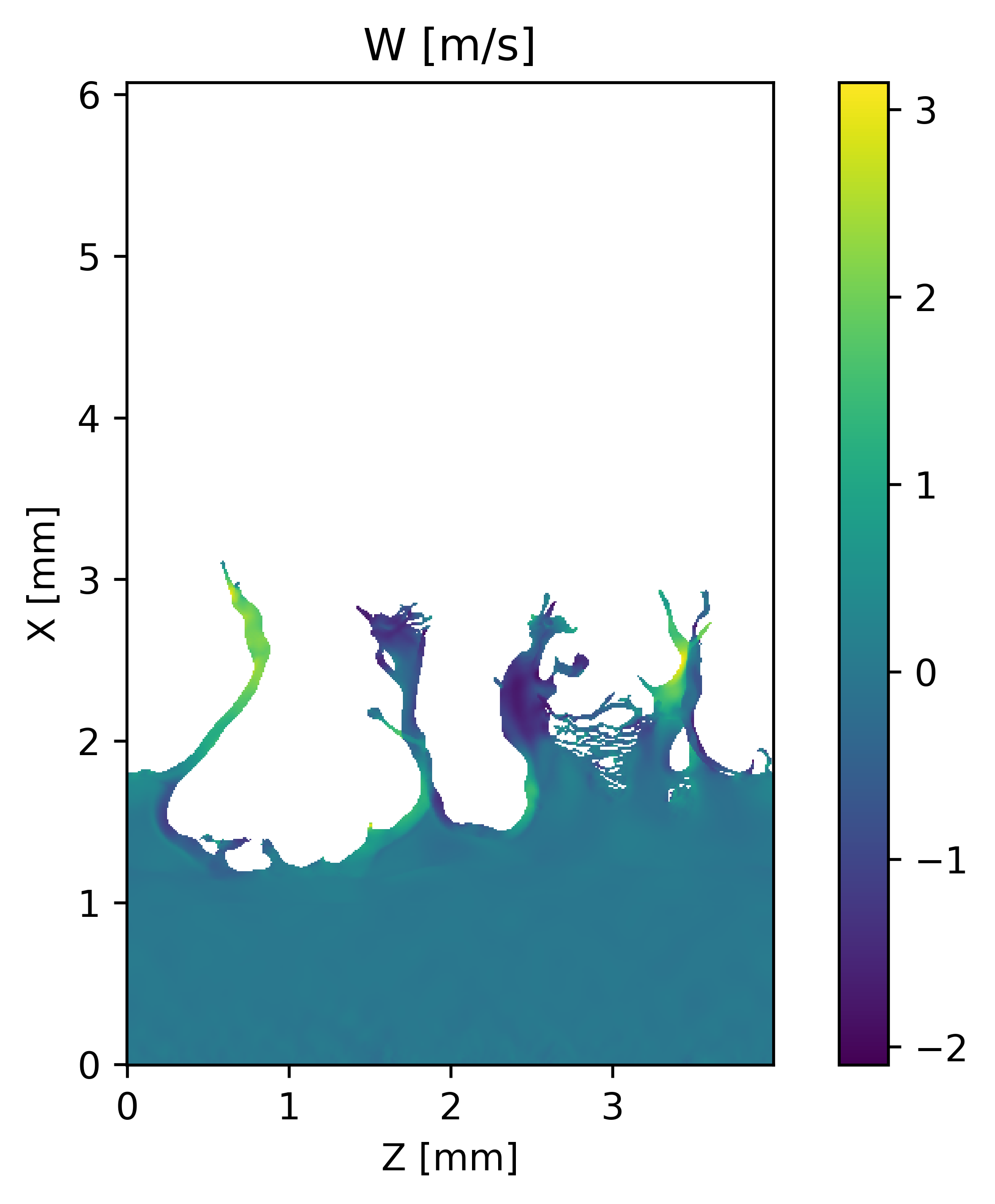}
%	\centering
	\caption{Liquid core velocity components projected on the XZ plane for a given time step of the $h_l \neq f(t,z)$ case.}
	\label{fig:pospro_vels_projection}
\end{figure}

In this work, ligament length data are complemented with ligament velocity data. During the 2D ligament projection step, at each XZ plane location, liquid velocities are obtained by adding the values of velocity along the Y axis weighted by $\alpha$. An example of the resulting velocity projected data for a given time step is shown in Figure \ref{fig:pospro_vels_projection}). Velocity data at the tip of each ligament $i$ ($\vb{u_{str2D,i}}$) are obtained from this projection.

Besides, a characteristic streamwise velocity of the liquid film is also computed according to the procedure by Warncke et al. \cite{Warncke2017}. From the liquid volume projection (Figure \ref{fig:pospro_ligs_detection} Step 1) of a given time step $t_k$, the furthest location of the attached liquid ($x_{max,zk}$) is determined for each $z$ position. The streamwise deformation velocity at each $z$ location is directly computed among time steps $t_k$ and $t_{k-1}$ as $u_{def,zk} = ( x_{max,zk} - x_{max,zk-1} ) / \Delta t$. The characteristic film deformation velocity of the $t_k$ time step ($u_{def,k}$) is then defined by the 90\% percentile of all $u_{def,zk}$ locations.

\begin{figure}[htbp]
	\centering
	\includegraphics[width=0.7\textwidth]{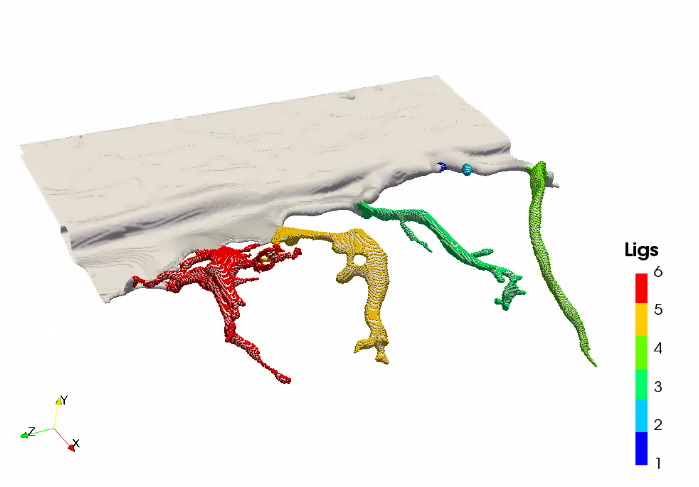}
%	\centering
	\caption{Sample of ligament detection in the 3D domain for a given time step of the $h_l \neq f(t,z)$ case.}
	\label{fig:pospro_ligs3d_measure}
\end{figure}

After 2D data for validation are obtained, a novel 3D methodology is proposed to prevent DNS information loss during the projection step. The objective is to preserve the 3D nature of the ligaments both in terms of ligament size and tip velocity components for a more realistic description of these structures. To this end, the cutting mask obtained in the 2D analysis is extruded along the Y axis so it can be used as a 3D cutting mask. When applied to the connected liquid in the domain, only the ligaments are left. It is then possible to apply another connectivity filter, identifying ligaments sizes and positions as depicted in Figure \ref{fig:pospro_ligs3d_measure}. Properly locating ligaments and the film in the 3D domain in this way then allows processing string tips, string lengths $L_{str3D,i}$ with the Fast Marching Method (Method 2), and the velocities $\vb{u_{str3D,i}}$ directly from the DNS output data.

Regardless of the geometrical treatment given to the ligaments (2D and 3D), some additional quantities derived from their features can be used to globally characterize the breakup process. A global breakup length is defined by averaging ligament lengths for all time steps ($L_{bu} = \overline{L_{str}}$) and a mean ligament tip velocity can be computed by averaging ligament velocities for all time steps ($\vb{u_{lig}} = \overline{\vb{u_{str}}}$). Besides, a mean film deformation speed is also computed as the average among time steps ($u_{def} = \overline{u_{def,k}}$). From these quantities, a mean breakup frequency ($f_{bu}$) is estimated according to Eq. (\ref{eq:fbu_ligament}) \cite{Warncke2017}:
\begin{equation}
    f_{bu} = \frac{u_{def}}{L_{bu}}
    \label{eq:fbu_ligament}
\end{equation}

%%% DROPLETS
\begin{itemize}
    \item Droplet cloud analysis
\end{itemize}

Focusing on the droplet cloud, it has been processed through the same method used by Crialesi-Esposito et al. and validated for a round spray \cite{CrialesiEsposito2022}, which is here summarized. Each individual droplet $i$ is assigned its volumetric diameter according to Eq. (\ref{eq:dv_droplet}), taking the assumption of spherical droplets. 
\begin{equation}
d_{V,i} = \sqrt[3]{\frac{6}{\pi}V_{drop,i}}
\label{eq:dv_droplet}
\end{equation}

where the droplet $i$ liquid volume ($V_{l,drop,i}$) is calculated as the sum of the liquid volume for every cell $j$ belonging to the droplet, as given by Eq. (\ref{eq:vl_droplet}):
\begin{equation}
 V_{drop,i} = \sum_{j=1}^{Ncells} {\alpha}_j {V_{cell,j}}
\label{eq:vl_droplet}
\end{equation}

where $Ncells$ is the number of cells representing droplet $i$. Droplet velocities are computed as weighted averages of the velocity from each cell $j$ composing the droplet as shown in Eq. (\ref{eq:u_droplet}), using the liquid volume fraction as the weighing factor:
\begin{equation}
 \vb{u_{drop,i}} = \frac{\sum_{j=1}^{Ncells} \alpha_j \cdot \vb{u_j}}{\sum_{j=1}^{Ncells} \alpha_j}
\label{eq:u_droplet}
\end{equation}

Any other droplet characteristics of interest (local $Re$, local $We$, etc.) can be computed from these magnitudes. Probability density function (PDF) graphs of both diameters and velocities are provided and analyzed in Section \ref{subsec:results_drops}. Droplets with $d_v < \SI{20}{\micro\meter}$ are discarded from the analysis for two reasons: on the one hand, this is the smallest droplet size detected by the reference experiments used for validation \cite{Warncke2017}. On the other hand, this value corresponds to twice the cell size realized by the proposed VOF-DNS simulations.

\section{Validation and results} \label{sec:results}
Results of the methodology are presented in this Section. First, each of the precursor LES is independently validated against experimental data. Next, results from the atomizing edge DNS for the two computed cases are presented, both from a qualitative and a qualitative standpoint. Quantitative results are also compared to experimental data for validation purposes.

\subsection{Validation of the precursor simulations}

\subsubsection{Single-phase flow LES}
\begin{figure}[htbp]
	\centering
	\includegraphics[width=1.0\textwidth]{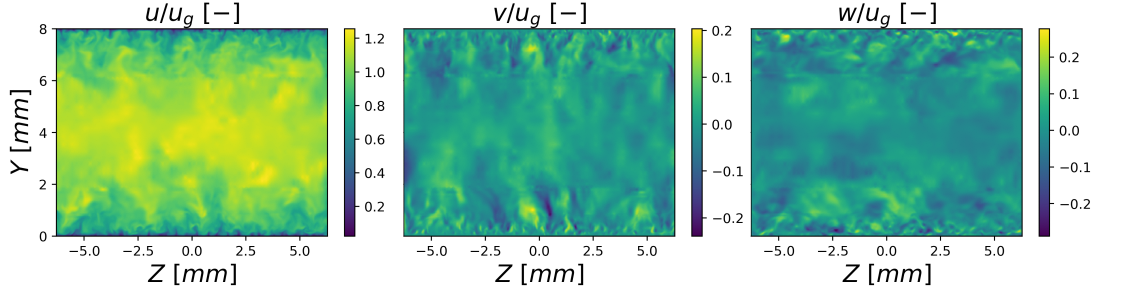}
	\centering
	\caption{Snapshots with the three normalized components of the velocity at $t = \SI{2.5}{\milli\second}$. Note the different scale in each case.}
	\label{fig:validation_1phase_UplanesYZ}
\end{figure}

For illustrating purposes, Figure \ref{fig:validation_1phase_UplanesYZ} shows an example of the instantaneous data samples in a YZ plane mentioned in Section \ref{sec:methodology}, where the spatial variability of each velocity component can be appreciated.

\begin{figure}[htbp]
	\centering
	\includegraphics[width=0.45\textwidth]{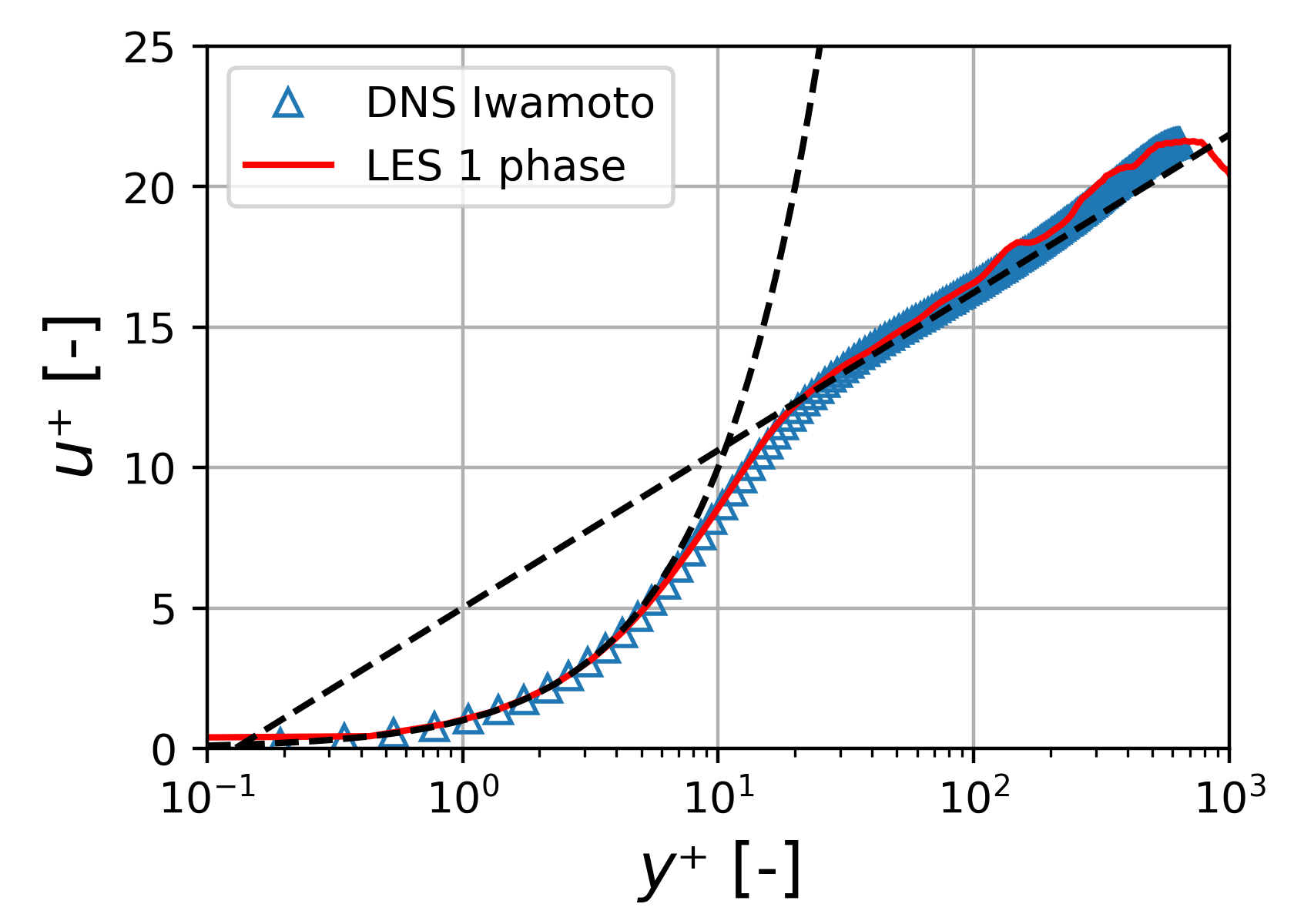}
	\includegraphics[width=0.45\textwidth]{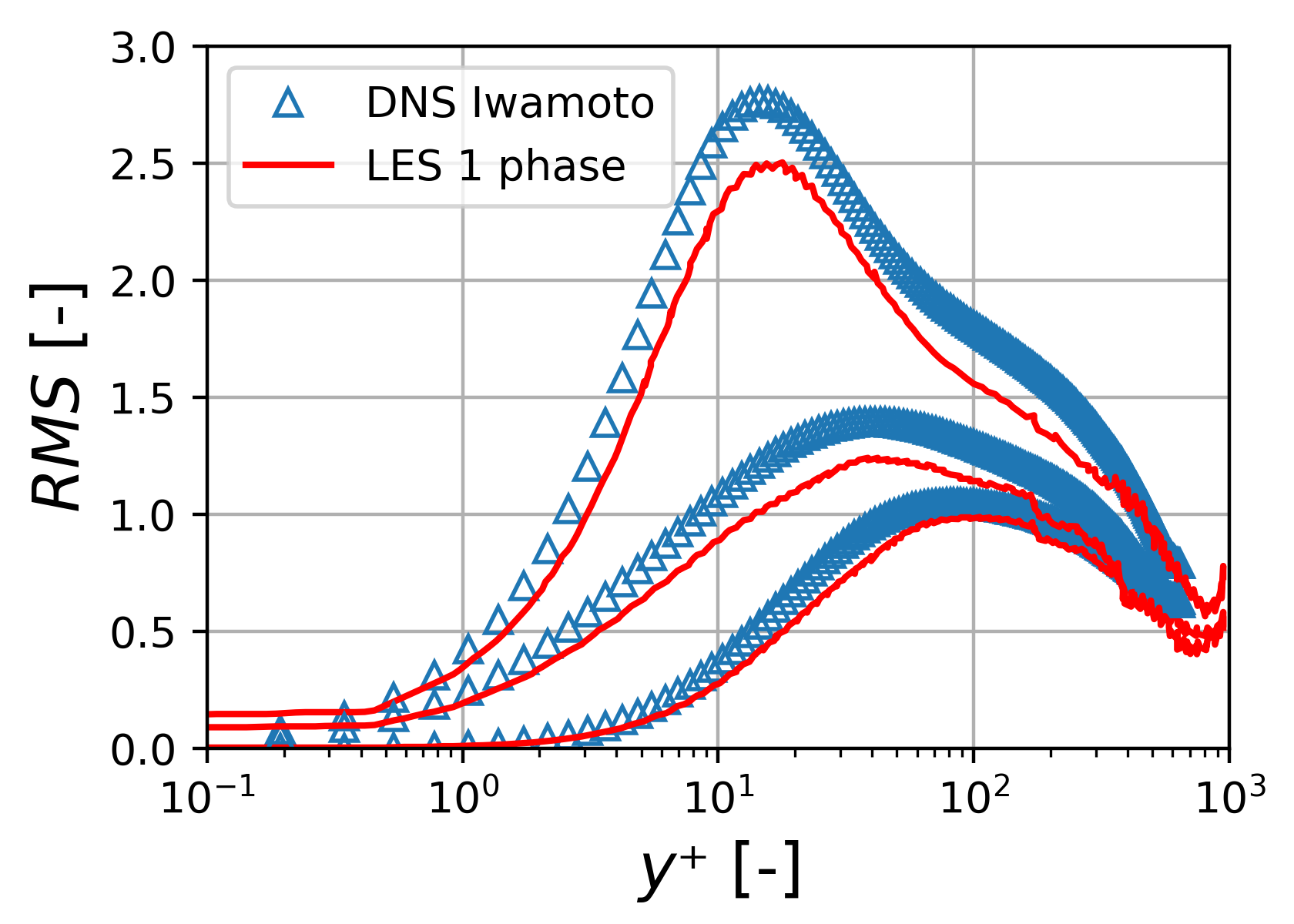}
	\centering
	\caption{Comparison of the single-phase LES temporally and spatially averaged velocity results against DNS data from Iwamoto et al. \cite{Iwamoto2002}: non-dimensional mean streamwise velocity profile (left, with the theoretical law of the wall sublayers also depicted in black), non-dimensional root mean square velocity components (right).}
	\label{fig:validation_1phase_yRMS}
\end{figure}

The velocities of these YZ planes are spatially averaged over the Z axis and temporally averaged for a large window in order to analyze the wall-normal velocity profile and whether it conforms with the law of the wall. Results for the mean dimensionless streamwise velocity profile are shown in Figure \ref{fig:validation_1phase_yRMS} (left), where they are also compared to the theoretical law of the wall and to accepted turbulent channel flow DNS data by Iwamoto et al. \cite{Iwamoto2002} for $Re_{\tau} \approx 650$. LES data show a reasonable agreement with the theoretical relations, with a smooth transition from the viscous sublayer to the log-law region. The agreement is comparable to that of a DNS computation, highlighting the level of resolution achieved by this precursor LES.

The root mean square (RMS) of each velocity component are similary computed and averaged. Their wall-normal evolution is plotted against Iwamoto's DNS data in Figure \ref{fig:validation_1phase_yRMS} with a similar level of agreement, although a slight underprediction is present for all components.

Additionally, the turbulence spectra were also analyzed through the autocorrelations of the velocity components to ensure sampled data correspond to a statistically steady time series and no artificially induced frequencies are passed on to subsequent simulations.

\subsubsection{Two-phase flow LES}

\begin{figure}[htbp]
	\centering
	\includegraphics[width=1\textwidth]{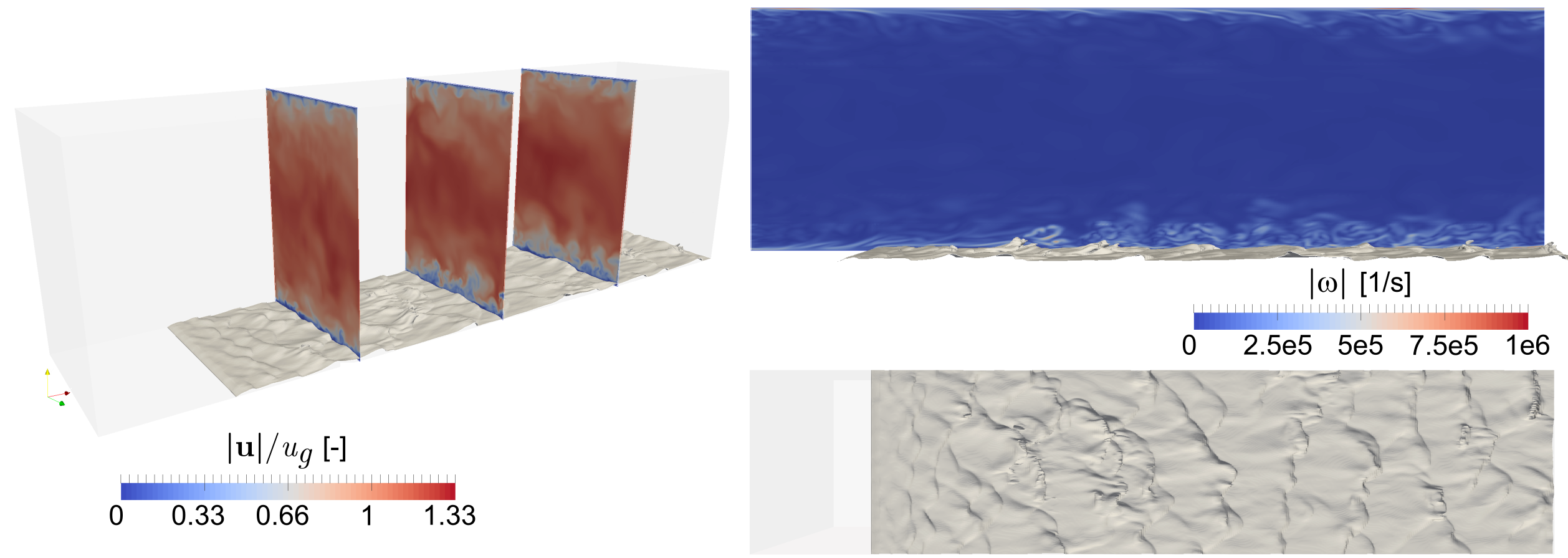}
	\centering
	\caption{Qualitative appearance of the liquid film on the prefilmer depicted through a $\alpha = 0.5$ iso-surface for a given time instant. 3D representation with velocity data at three different streamwise locations (left), side view with vorticity magnitude in a mid-plane (right, top), top view (right, bottom).}
	\label{fig:results_2phaseLES}
\end{figure}

In the case of the two-phase flow LES, results were analyzed once the liquid injection transient was finished. Figure \ref{fig:results_2phaseLES} (top) shows the qualitative appearance of the film for a given instant. Normalized velocity data at three different YZ planes are shown to illustrate the influence of the liquid film on the gaseous phase velocity, as a liquid-gas boundary layer is developed. Figure \ref{fig:results_2phaseLES} (right, top) depicts a side view of the liquid film. Vorticity magnitude data at the mid XY plane is shown, where it may be seen how the liquid-gas interaction along the film crests and valleys generates additional vortex to the ones generated at the gas-wall boundary layer on the upper wall. These vortex also spread for a larger distance from the film in the wall-normal direction than they do in the case of the gas-wall interaction. This fact also highlights the generation of a thicker boundary layer among liquid and gas. A top view of the film is also shown in Figure \ref{fig:results_2phaseLES} (bottom) to complete the qualitative appearance of the liquid film, showing the wavy behavior with crests and valleys being spahed and propagated in the streamwise direction. Several wave crests or valleys can be contained in the spanwise direction.

\begin{figure}[htbp]
	\centering
	\includegraphics[width=0.6\textwidth]{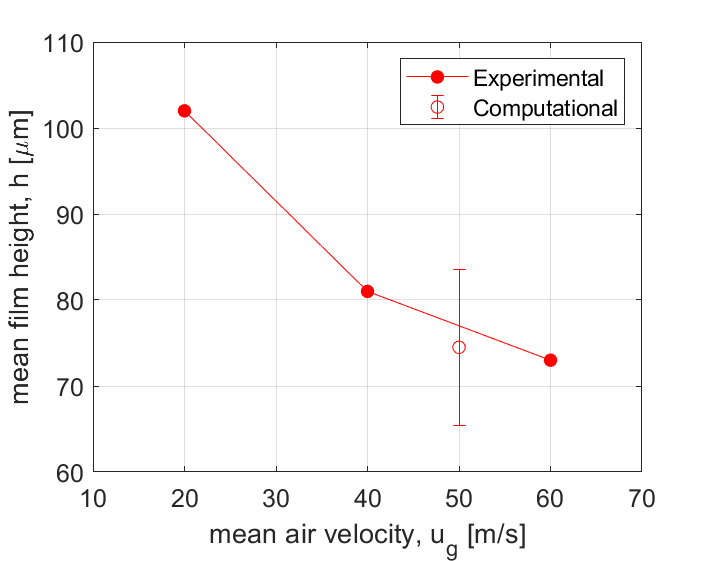}
	\centering
	\caption{Mean film height as a function of the mean gas velocity. The tested operating condition is plotted together with experimental data from Gepperth et al. \cite{Gepperth2010} for Shellsol D70 and the same normalized volumetric flow rate.}
	\label{fig:results_2phaseLES_filmHeight}
\end{figure}

In this case, liquid volume fraction ($\alpha$) data sampled at the three YZ planes and at the domain mid-plane (XZ) were processed to compute the liquid film height at all locations and their temporal evolution. The analysis allowed determining that data at the $x = 2 \cdot h_c$ location is already not influenced by the inlet boundaries and thus can be passed on to the atomizing edge DNS. The temporally averaged mean liquid film height and the corresponding standard deviation found at this location are plotted in Figure \ref{fig:results_2phaseLES_filmHeight} against experimental data for similar conditions reported by Gepperth et al. \cite{Gepperth2010}. Further averaged results at this location are displayed in Table \ref{tab:results_2phaseLES}. It must be stated that the mean temporal frequency value obtained by the LES ($\overline{f_{film}} = \SI{585.9}{\hertz}$) is also very close to the one reported by Holz et al. \cite{Holz2018} ($\overline{f_{film}} = \SI{595}{\hertz}$) for the same operating condition. The predictions by this precursor 2-phase flow LES are validated by these means. The averaged value of $h_l$ is then passed on to the $h_l \neq f(t,z)$ atomizing edge DNS, whereas sampled data are provided to the $h_l = f(t,z)$ case.

\begin{table}[ht]
\caption{Temporally averaged data for the 2-phase flow LES at $x = 2 \cdot h_c$.}
\centering
\begin{tabular}{llS[table-format=3.1]r} 
\hline
Description & Parameter & {Value} & Units \\
\hline
Mean film height & $\overline{h_l}$ & 74.3 & $\mu$m \\ %\hline
Film height std deviation & $\sigma_{hl}$ & 7.5 & $\mu$m \\
Mean film wave frequency & $\overline{f_{film}}$ & 585.9 & Hz \\
Mean film wavelength & $\overline{\lambda_{film}}$ & 2.11 & mm \\
\hline
\end{tabular}
\label{tab:results_2phaseLES}
\end{table}

\subsection{Two-phase flow DNS simulations} %\label{sec:results_quali}

Results of the atomizing edge DNS are analyzed in the present Section. First, a qualitative analysis is carried out, describing the atomization mechanism and comparing its appearance for both tested strategies. Next, a quantitative analysis of the ligament development and the generated droplet cloud is carried out in Subsections \ref{subsec:results_ligs} and \ref{subsec:results_drops}, respectively.

\begin{figure}[htbp]
	\centering
	\includegraphics[width=1.1\textwidth]{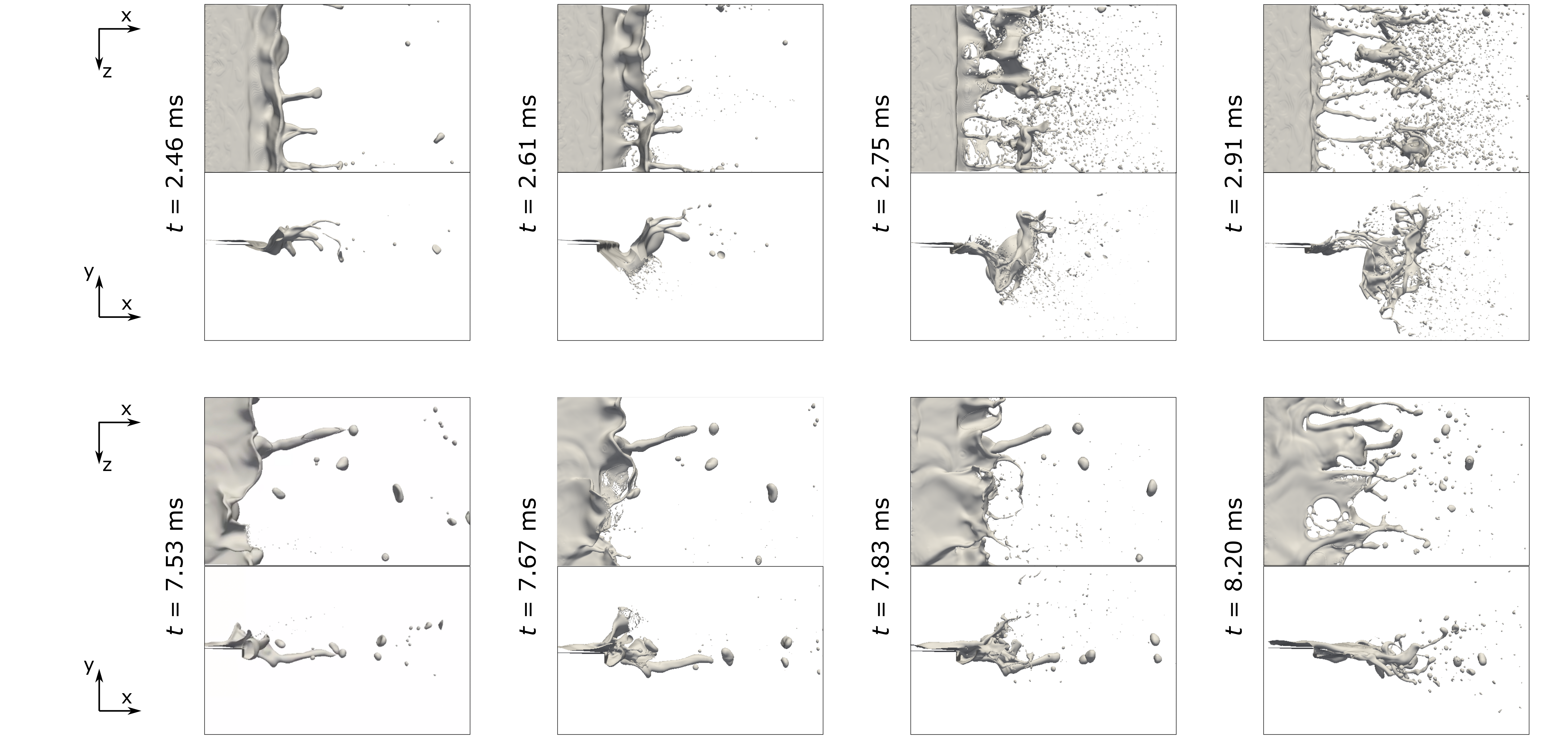}
	\centering
	\caption{Time sequence of the DNS results for the $h_l \neq f(t,z)$ case (top) and the $h_l \neq f(t,z)$ case (bottom). The liquid is depicted through an iso-surface of $\alpha = 0.5$.}
	\label{fig:results_DNS_qualitative}
\end{figure}

%%% Qualitative results
Figure \ref{fig:results_DNS_qualitative} illustrates the temporal evolution of the flow for a given breakup event of both simulated cases. To this end, it represents a series of snapshots depicting the evolution of the airblasted sheet. The liquid film and structures are characterized by iso-surfaces of $\alpha = 0.5$, as it is representative of the liquid-gas interface. Focusing on the $h_l \neq f(t,z)$ case on the one hand, the breakup mechanism phenomenology already described in the literature \cite{Gepperth2012, Chaussonnet2020, Chaussonnet2016, Sauer2016,  Warncke2017} is reproduced. First, the liquid transported along the prefilmer tends to accumulate behind the prefilmer edge, generating a liquid reservoir ($t = \SI{2.46}{\milli\second}$). Two structures then emanate from this reservoir: elongated structures (ligaments) on the one hand, and most importantly a liquid sheet that tends to flap upwards or downwards. This liquid sheet then gets exposed to the free gas stream coming from above or below the prefilmer surface, shaping bag-like structures (observed at $t = \SI{2.46}{\milli\second}$ and $t = \SI{2.61}{\milli\second}$) that keep growing in size. Eventually, surface tension cannot keep the bag structures attached to the liquid sheet, so they get punctured and break into droplets in the so-called \textit{bag breakup} ($t = \SI{2.61}{\milli\second}$ and $t = \SI{2.75}{\milli\second}$). Additionally, some rims are shaped from the sides of the bag structures ($t = \SI{2.75}{\milli\second}$), generating separated ligaments that also break into droplets ($t = \SI{2.91}{\milli\second}$). Some of these ligaments keep attached to the liquid film that starts generating a new reservoir behind the prefilmer edge. If these ligaments do not detach among main breakup events, they get thicker by receiving some liquid transported from the film (back to $t = \SI{2.46}{\milli\second}$). Therefore, in the end, droplets are generated by two main mechanisms: bag breakup, which generates a large amount of small droplets only during the main breakup events; and ligament breakup, which generates a lower amount of relatively bigger droplets, some of them still being generated among breakup events. The coexistence of these distinct breakup mechanisms and the importance of the liquid accumulation process is justified considering the value of the momentum flux ratio ($M = 15.7$) achieved for this operating condition: Fern\'andez et al. \cite{Fernandez2009} found this value led to the so-called \textit{torn sheet breakup}. According to these authors, values of $M > 20$ would have led to the \textit{membrane sheet breakup}, with a low importance of the accumulation mechanism in favor of a direct disintegration in droplets behind the prefilmer edge; whereas $M < 5$ would have led to the \textit{stretched ligament breakup}, with a higher presence of ligaments generated from rims after bag breakup.

On the other hand, a look at the $h_l = f(t,z)$ case highlights the qualitative differences when accounting for the liquid film history upstream of the prefilmer edge. In this case, the liquid film above the prefilmer surface appears more wrinkled than in the $h_l \neq f(t,z)$ case, as a consequence of the two-phase LES data mapping procedure. Therefore, the reservoir behind the prefilmer edge is less uniformly distributed spanwise than it was for the constant film thickness case, as the film crests and valleys do not reach the reservoir in a synchronized manner (recall Figure \ref{fig:results_2phaseLES}). As a result, by the time a bag is formed and flapping, not the complete prefilmer span is filled with enough liquid to shape bags. Hence, while the breakup mechanism remains the same as in the $h_l \neq f(t,z)$ case, the aforementioned fact results in a less violent main bag breakup event. The amount of generated droplets is substantially lower, and the resulting cone-like injection zone is narrower in the $h_l = f(t,z)$ case than in the $h_l \neq f(t,z)$ case. Also, the shaped ligaments tend to be thicker, thus resulting in larger droplets from ligament breakup. In short, even though there are still main breakup events, the atomization seems inhibited to a certain extent, but seems more continuous in time.

%%% Quantitative results

\subsubsection{Ligaments results} \label{subsec:results_ligs}

\begin{figure}[htbp]
	\centering
	\includegraphics[width=0.5\textwidth]{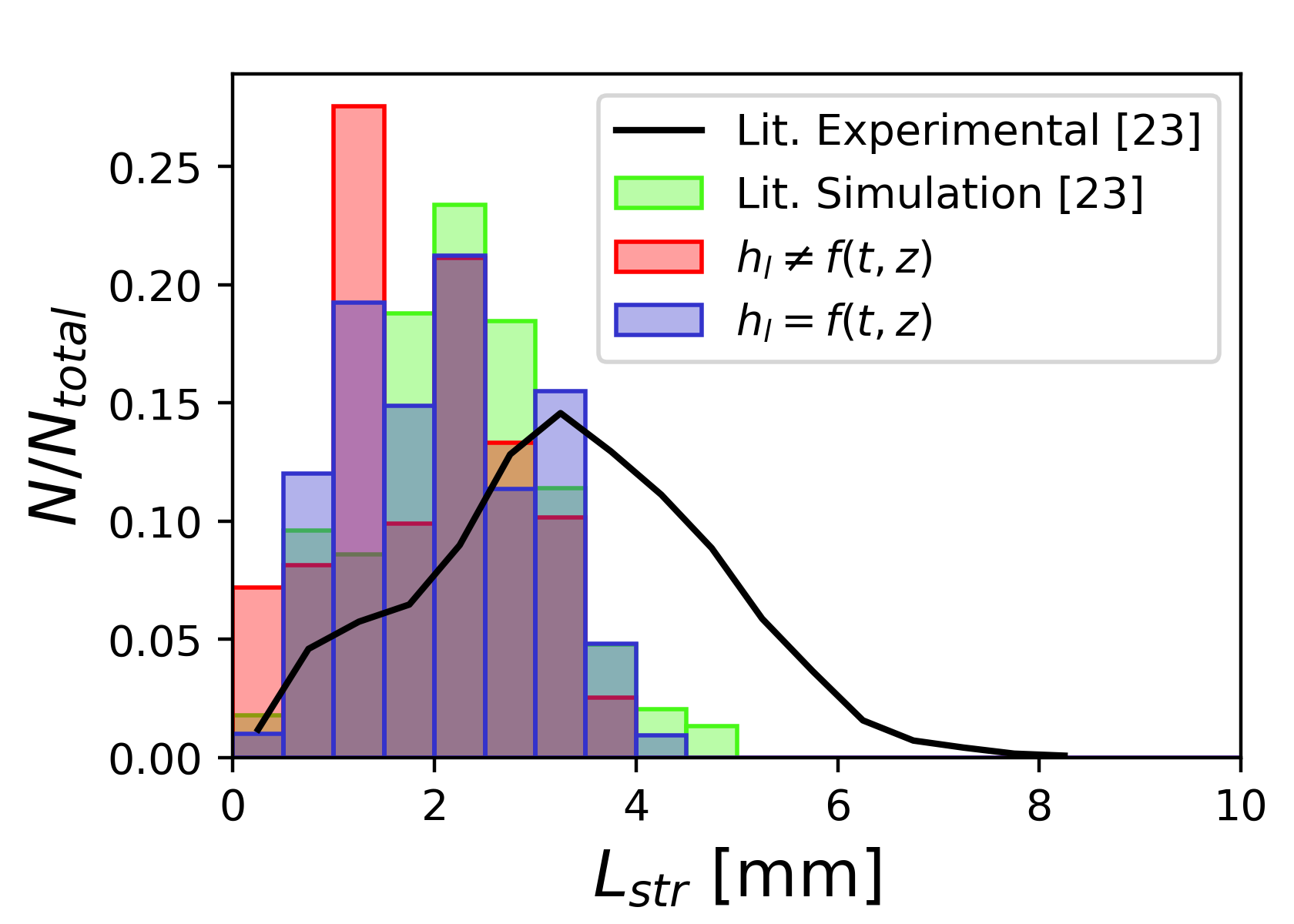}
%	\centering
	\caption{Ligament 2D length distribution for both simulated test cases (Method 1 used for ligament length measurement). Literature DNS and literature experiments are also reported \cite{Warncke2017}.}
	\label{fig:results_ligs2d_Lstr_validation}
\end{figure}

Ligaments are characterized in 2D through the liquid volume fraction $\alpha$ projection in the XZ plane and the identification of individual elongated structures, as detailed in Section \ref{sec:pospro}. The histogram of ligament lengths is shown in Figure \ref{fig:results_ligs2d_Lstr_validation} for both simulated cases and both experimental and DNS data from the literature. First, it must be mentioned that both ligament size distributions obtained by the DNS simulations presented in this investigation importantly differ from the experimental data reported in the literature. However, Warncke et al. \cite{Warncke2017} already justified that the experiment occasionally reports several ligament lengths beyond the DNS domain ($L_{str} > \SI{5}{\milli\meter}$), with a wide time window (i.e. a long amount of breakup events) being analyzed. In the simulations, no structure attached to the intact core was found to trespass the domain outlet, with maximum lengths around $L_{str} = \SI{4}{\milli\meter}$. Reliably predicting the maximum ligament length for the operating condition studied would then imply enlarging the DNS domain and analyzing a greater amount of breakup events. Both actions are beyond the current affordable computational resources for a DNS study. In any case, the comparison of both $h_l \neq f(t,z)$ and $h_l = f(t,z)$ cases with the literature DNS (which uses the same domain length as the one used in the present study) shows fairly similar distributions, with ligament lengths spanning from 0 to \SI{4.5}{\milli\meter} and most values being found among 1.5 and \SI{2.5}{\milli\meter}. An improvement in the predictions is found in the $h_l = f(t,z)$ case with respect to the $h_l \neq f(t,z)$ case, as the distribution as a whole and its mode are shifted towards longer lengths when the liquid film thickness evolution is accounted for at the DNS inlet. Besides, the distribution more closely resembles a normal distribution, as it was found for the experiments. The irregular shape from the distribution can be attributed to the fact that only 3 breakup events could be simulated.

\begin{figure}[htbp]
	\centering
	\includegraphics[width=0.48\textwidth]{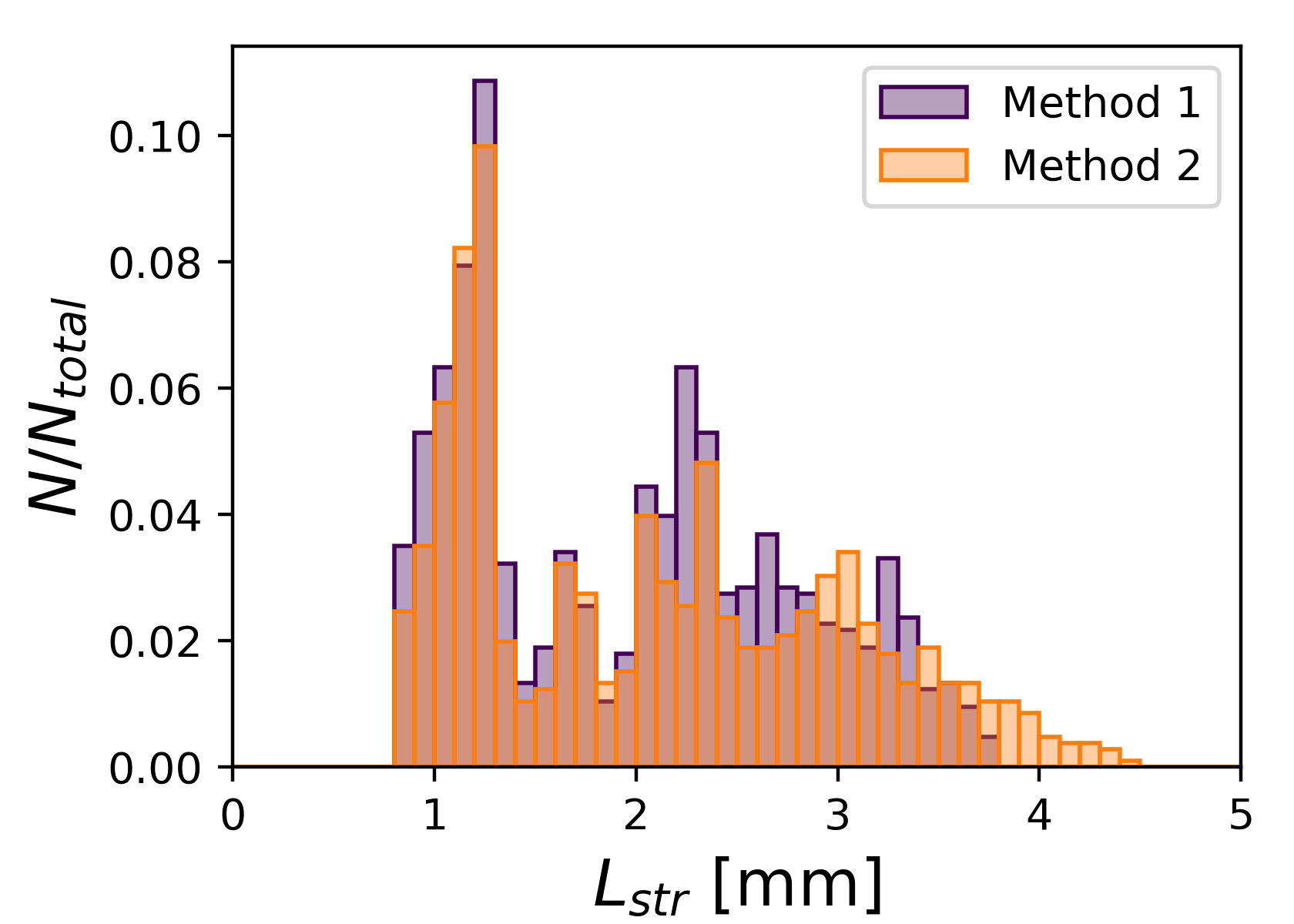}
	\includegraphics[width=0.46\textwidth]{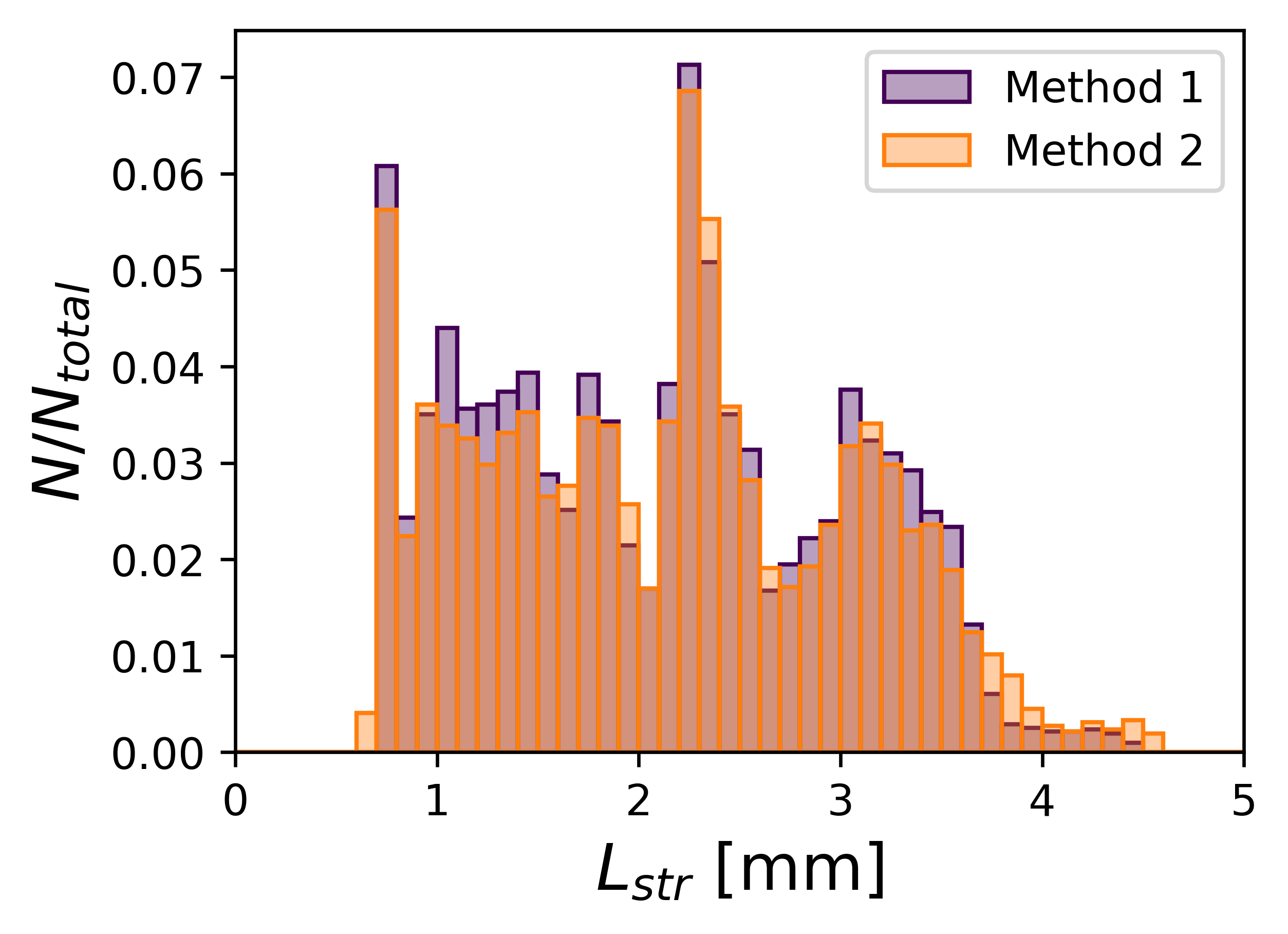}
%	\centering
	\caption{Comparison among the ligament 2D length distributions found for Method 1 and Method 2 for the $h_l \neq f(t,z)$ case (left) and $h_l = f(t,z)$ case (right).}
	\label{fig:results_ligs2d_Lstr_methods}
\end{figure}

Before analyzing the results found through the 3D post-processing strategy, it is interesting to assess the differences among Method 1 (axial distance method) and Method 2 (Fast Marching Method) for ligament length determination, since 3D sizes can only be determined through Method 2. Figure \ref{fig:results_ligs2d_Lstr_methods} shows this comparison for the $h_l \neq f(t,z)$ and $h_l = f(t,z)$ cases. In both cases, the distributions found through both methods are relatively similar. Ligament sizes predicted through Method 2 are statistically slightly displaced towards larger values, as expected considering that the whole ligament path is computed as opposed to only accounting for streamwise variations in the ligament tip location. Anyway, this method displays its potential for the 3D ligament detection case, for which it is strictly needed.

\begin{figure}[htbp]
	\centering
	\includegraphics[width=0.5\textwidth]{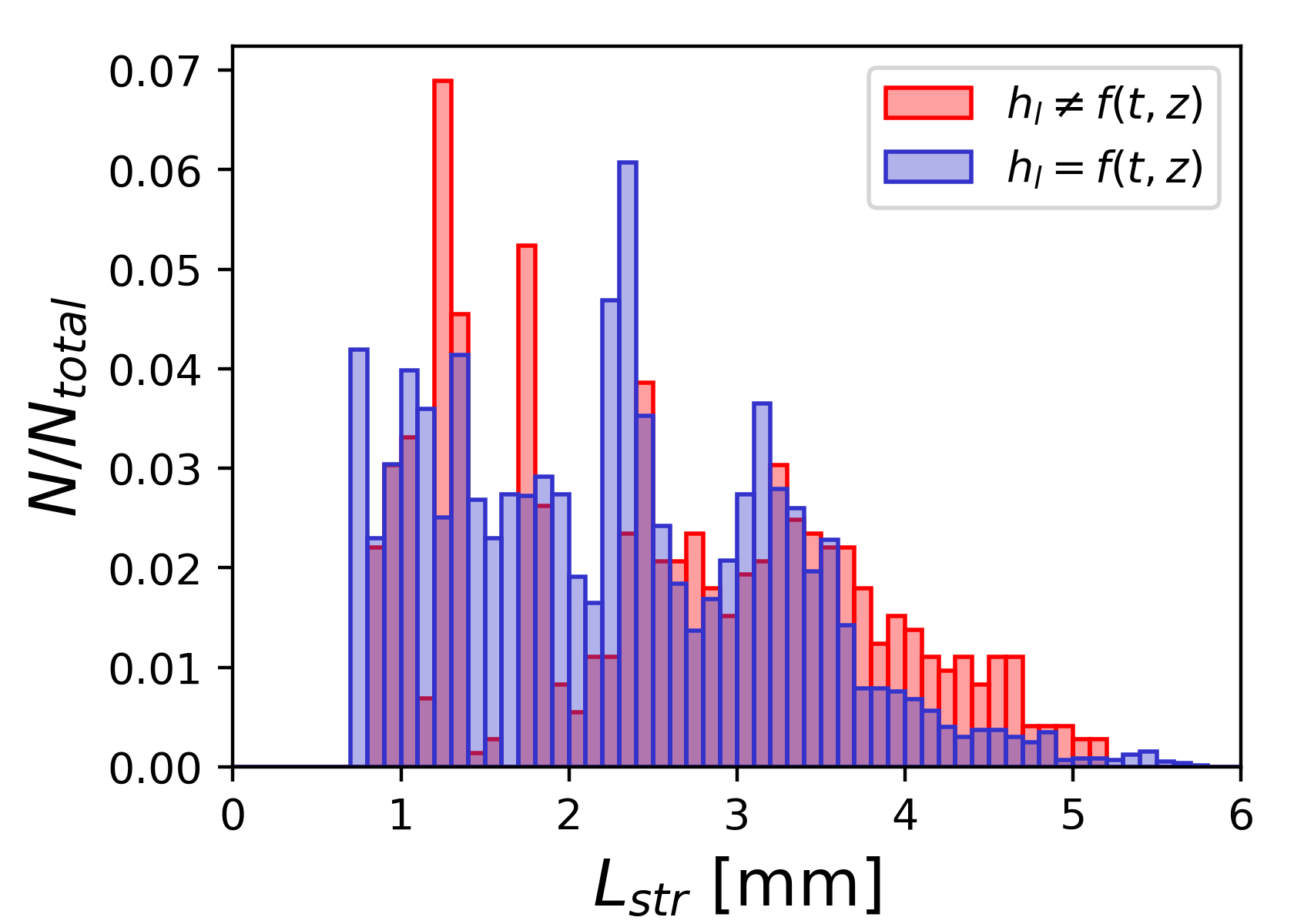}
%	\centering
	\caption{Ligament 3D length distributions found for both simulated test cases (Method 2 used for ligament length measurement).}
	\label{fig:results_ligs3d_Lstr_cases}
\end{figure}

Once the differences among ligament length determination methods have been assessed, Figure \ref{fig:results_ligs3d_Lstr_cases} depicts the results obtained in the ligament 3D lengths between simulated test cases. Focusing on one of the simulated test cases separately, a certain shift towards greater ligament lengths is observed compared to the analogous 2D distributions with Method 2 from Figure \ref{fig:results_ligs2d_Lstr_methods}, as logical. Except for the shortest ligaments (hardly distinguishable from the intact film), the $h_l \neq f(t,z)$ case more closely resembles a normal distribution. Despite the comparison in 2D (Figure \ref{fig:results_ligs2d_Lstr_validation}) showing longer ligament lengths for the $h_l = f(t,z)$ case, the 3D ligament length distribution found for the $h_l \neq f(t,z)$ case is more shifted towards greater values. Hence, these differences must be attributed to the more accused flapping observed when liquid accumulation behind the prefilmer edge is synchronized along the span, as commented in the view of Figure \ref{fig:results_DNS_qualitative}. This fact highlights that any ligament length comparisons with experimental data are not fully representative unless flapping is also taken into account.

\begin{figure}[htbp]
	\centering
	\includegraphics[width=0.42\textwidth]{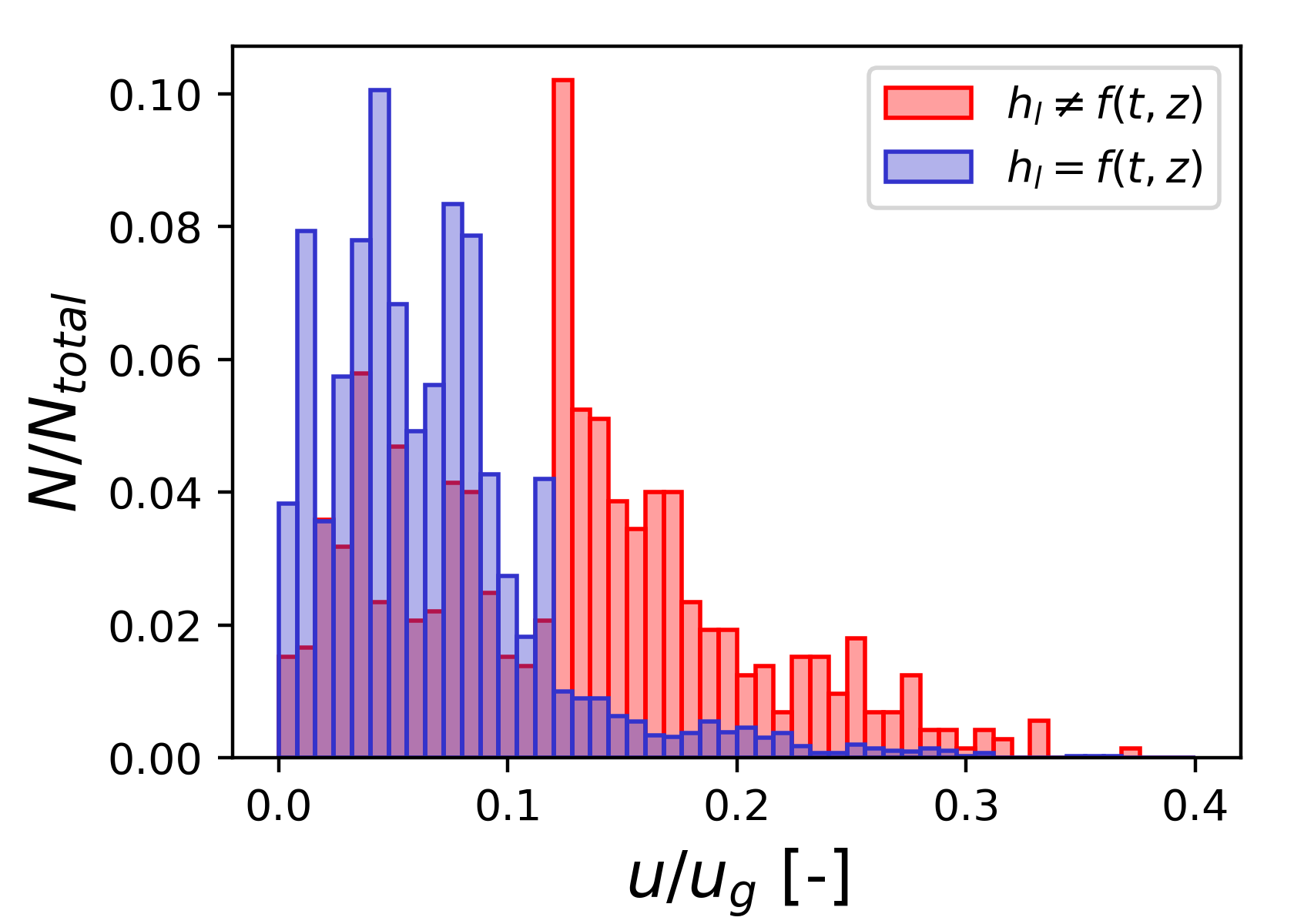}
	\includegraphics[width=0.42\textwidth]{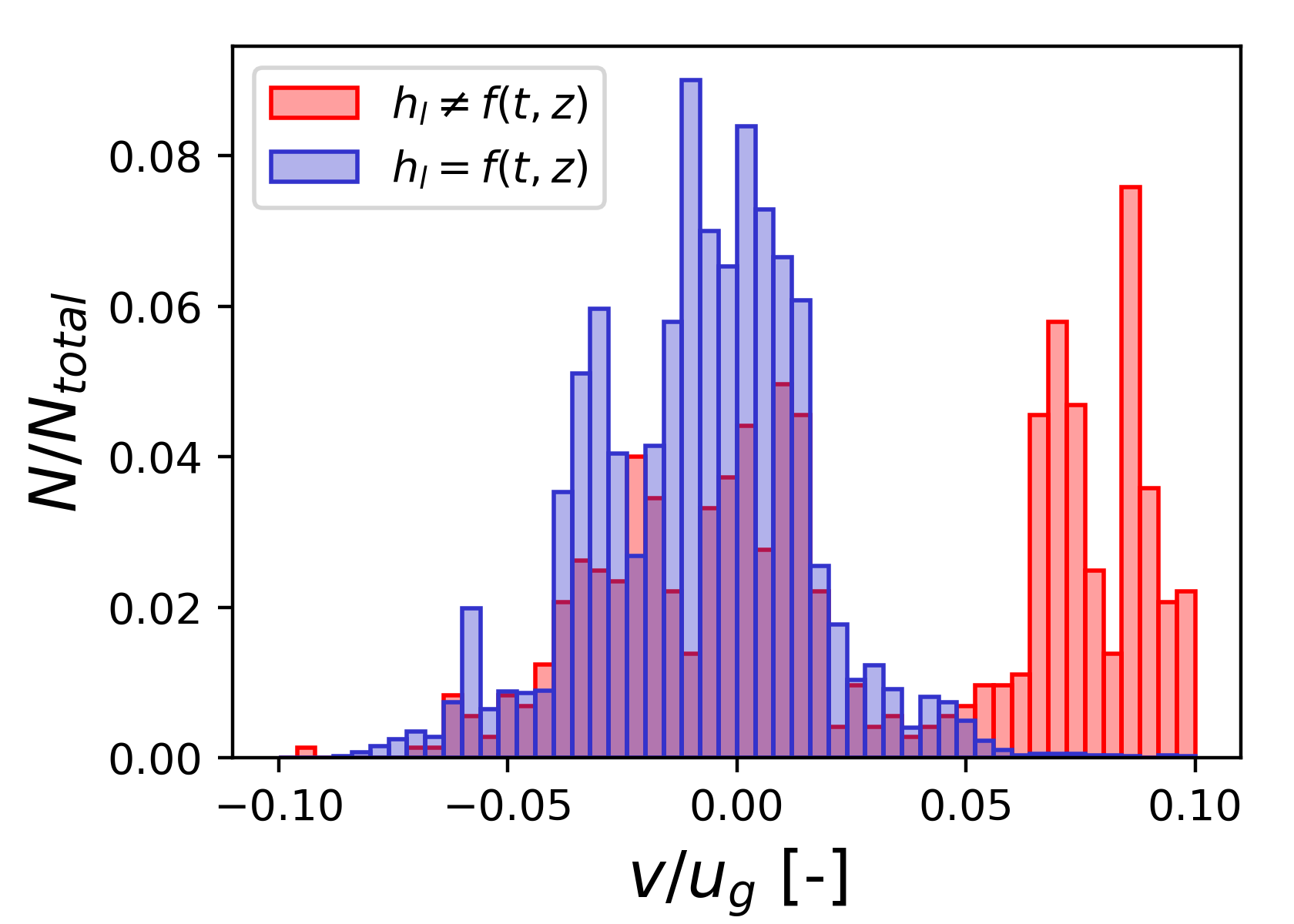}
	\includegraphics[width=0.42\textwidth]{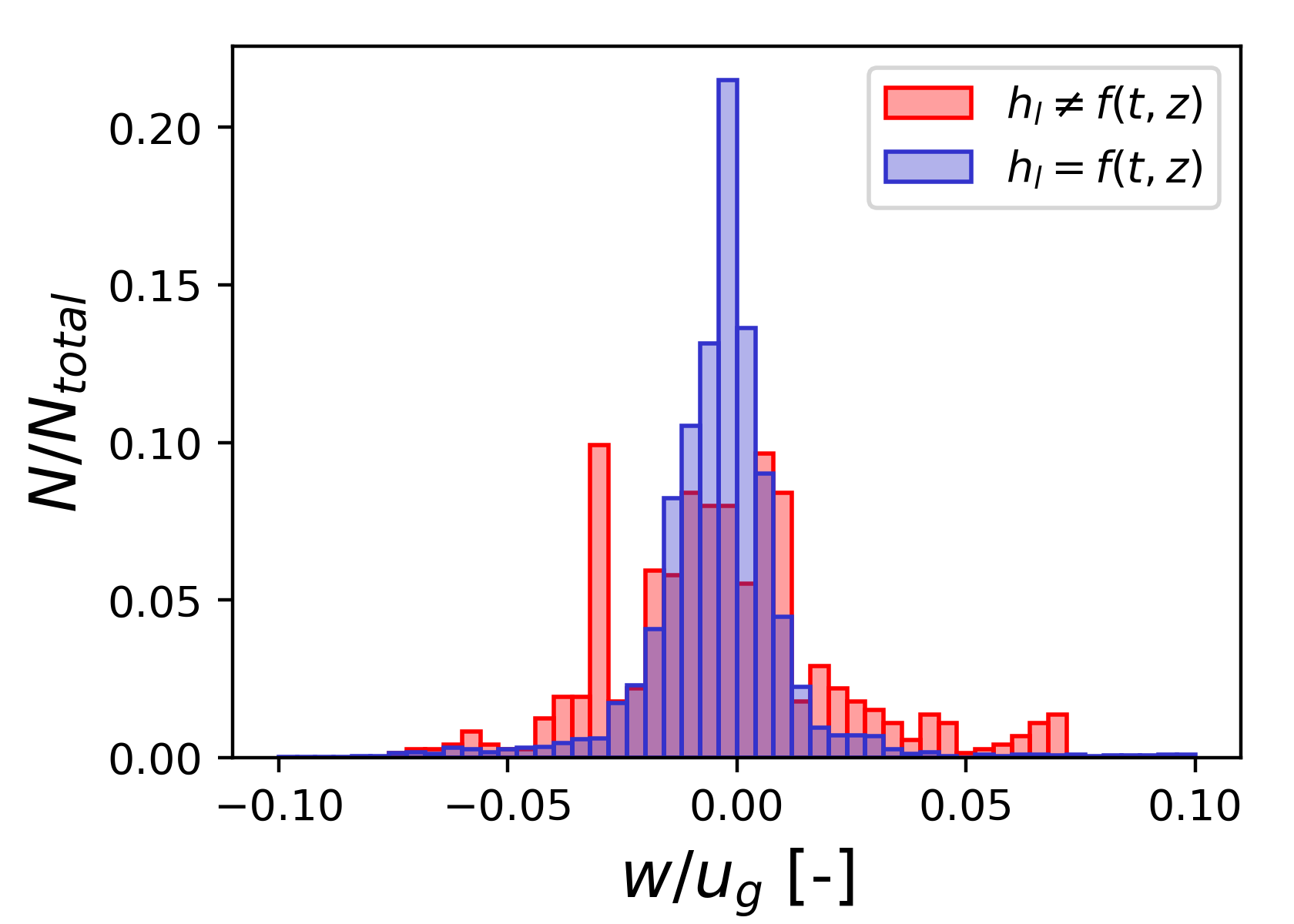}
%	\centering
	\caption{Distributions of 3D ligament tip velocity components for both simulated test cases: streamwise velocity (top left), wall-normal velocity (top right), spanwise velocity (bottom). All values are normalized with the gas bulk velocity.}
	\label{fig:results_ligs3d_ustr_cases}
\end{figure}

\begin{figure}[htbp]
	\centering
	\includegraphics[width=0.48\textwidth]{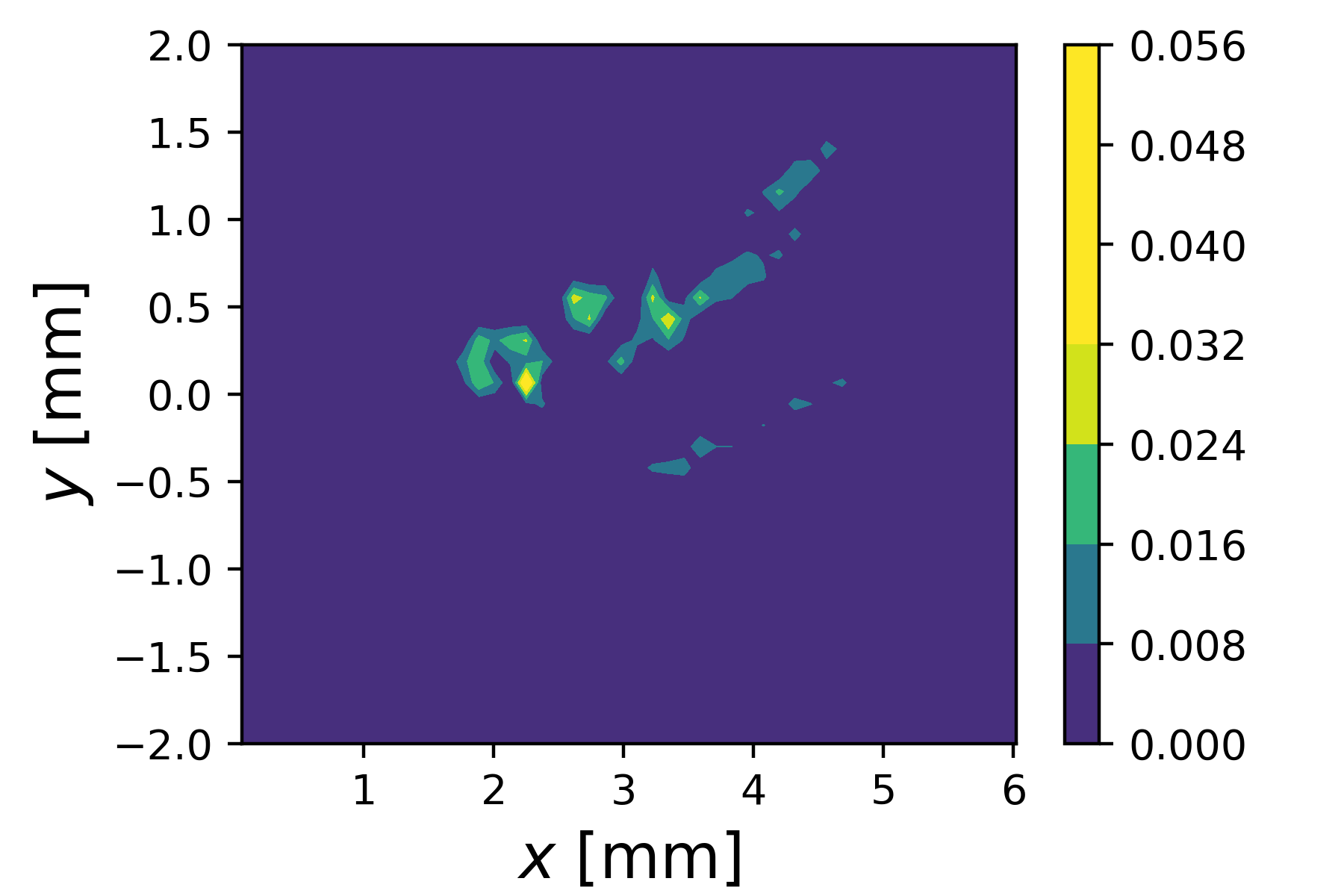}
	\includegraphics[width=0.48\textwidth]{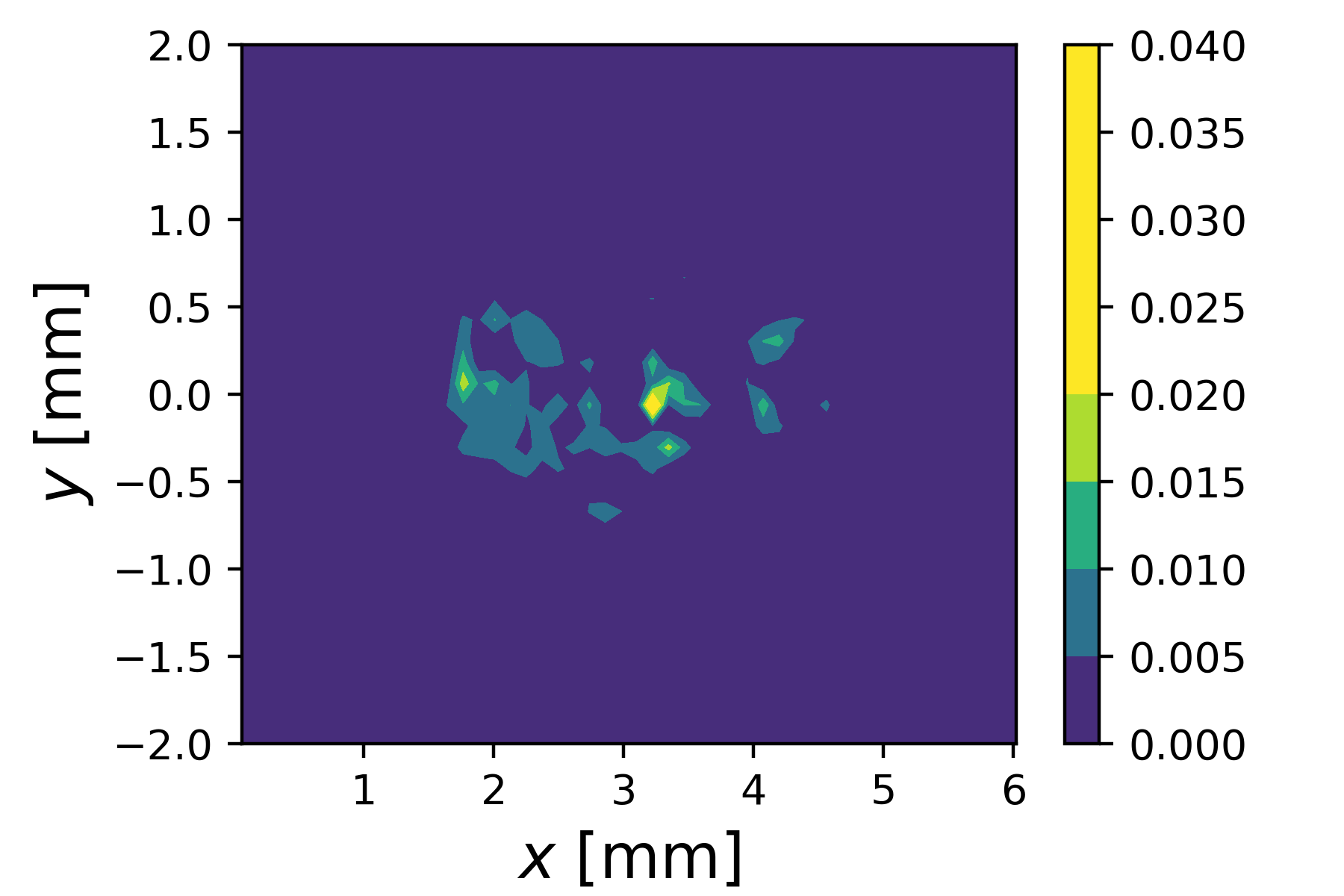}
%	\centering
	\caption{Contours displaying the probability of finding a ligament tip in the different locations of the XY plane: $h_l \neq f(t,z)$ case (left) and $h_l = f(t,z)$ case (right).}
	\label{fig:results_ligs3d_locations}
\end{figure}

As far as ligament tip velocities $\vb{u_{str}}$ are concerned, the distributions obtained for each of the three components in each simulated case are shown in Figure \ref{fig:results_ligs3d_ustr_cases}. Important differences are found in all cases. Focusing on the streamwise velocity component, the distribution for the $h_l \neq f(t,z)$ case is significantly shifted towards higher values (the distribution is centered around $u/u_g =$ 0.12) than the distribution for the $h_l = f(t,z)$ case (most ligaments found in the $u/u_g =$ 0 to 0.1 range). This may be explained considering the locations at which it is more probable to find a ligament tip in each simulated case, shown in Figure \ref{fig:results_ligs3d_locations}. In the $h_l \neq f(t,z)$ case, the liquid film flaps more importantly about the prefilmer center line (recall also Figure \ref{fig:results_DNS_qualitative}). Thus, in this case the ligaments extend towards more external wall-normal locations, where the influence from the liquid and gas-liquid boundary layers is less accused. At these locations the gaseous phase travels faster, leading to a greater momentum exchange with the liquid film and the resulting ligaments, increasing their axial velocities. The locations at which the ligament tips are found also explain the different distributions obtained in Figure \ref{fig:results_ligs3d_ustr_cases} for the wall-normal ligament tip velocity components: the $h_l = f(t,z)$ case resembles a normal distribution centered about $v/u_g =$ 0, as the ligament tips are equally found around the prefilmer center line in Figure \ref{fig:results_ligs3d_locations}. However, the $h_l \neq f(t,z)$ case yields a bimodal distribution, one of the modes being centered around a value $v/u_g \geq 0$. This mode corresponds to the fact that, in this case, the ligament tips are preferentially found above the prefilmer edge (see Figure \ref{fig:results_ligs3d_locations}) or close to its center (please note that this result might be biased by the fact that only 3 breakup events could be simulated, potentially not resulting in a statistically representative sample in this strategy). Ligaments that extend above the prefilmer tend to move away from the prefilmer surface in the wall-normal direction as they penetrate axially, thus possessing positive wall-normal velocities. Finally, the differences among test cases in the spanwise component distributions are less accused, being centered around $w/u_g = 0$ in both cases. The deviation about this value is more accused in the $h_l \neq f(t,z)$ case, since the film disintegration into ligaments is more violent in this simulation.

Characteristic quantities from the ligament analysis are displayed in Table \ref{tab:results_ligs}. As expected according to the view of the size PDFs, the mean breakup length $L_{bu}$ is sensibly underpredicted in all cases, but it reaches values similar to the literature DNS \cite{Warncke2017, Mukundan2022}. The 2D analysis used for validation shows that the breakup length prediction for the $h_l = f(t,z)$ case is slightly closer to the literature data. This fact, together with a better prediction of $u_{def}$, leads to a better agreement in mean breakup frequency $f_{bu}$ than the one obtained through the $h_l \neq f(t,z)$ case, although the value is overpredicted compared to the experiments due to the large underprediction in breakup length. In any case, it is observed that mean values may not be representative of the agreement in the related distributions. The values extracted through the developed 3D analysis show that the breakup length is more noticeably increased from the 2D values in the $h_l \neq f(t,z)$ case than in the $h_l = f(t,z)$ case, due to the larger extension of the ligaments in the wall-normal direction found for this case. The film deformation velocity is however reduced, thus leading to lower breakup frequencies if they are computed from the 3D analysis.

\begin{table}[ht]
\caption{Characteristic quantities obtained from the ligament 2D and 3D analysis for both simulations and size post-processing methods. Literature DNS and literature experimental data from \cite{Warncke2017} are also provided. Note: the $u_{lig}$ value of the literature experimental data was found for $\dot{V}/b = 25$ mm\textsuperscript{2}/s.}
\centering
\begin{tabular}{lS[table-format=1.1]S[table-format=1.2]S[table-format=2.2]S[table-format=1.2]} 
\hline
Source & {$L_{bu}$ [mm]} & {$u_{lig}$ [m/s]} & {$u_{def}$ [m/s]} & {$f_{bu}$ [kHz]} \\
\hline
Lit. Exp. 2D \cite{Gepperth2012} & 3.2 & 5.00 & \text{-} & 4.91  \\ %\hline
Lit. DNS 2D \cite{Warncke2017} & 2.2 & \text{-} & 8.2 & 3.73  \\ 
\hline
$h_l \neq f(t,z)$ 2D & 2.0 & 5.79 & 6.53 & 3.27  \\  
%\hline
$h_l = f(t,z)$ 2D & 2.1 & 6.2 & 11.97 & 5.70  \\ 
\hline
$h_l \neq f(t,z)$ 3D & 2.56 & 6.87 & 6.28 & 2.45 \\ 
$h_l = f(t,z)$ 3D & 2.25 & 3.50 & 11.51 & 5.12 \\ 
\hline
\end{tabular}
\label{tab:results_ligs}
\end{table}

\subsubsection{Droplet cloud results} \label{subsec:results_drops}

\begin{figure}[htbp]
	\centering
	\includegraphics[width=0.85\textwidth]{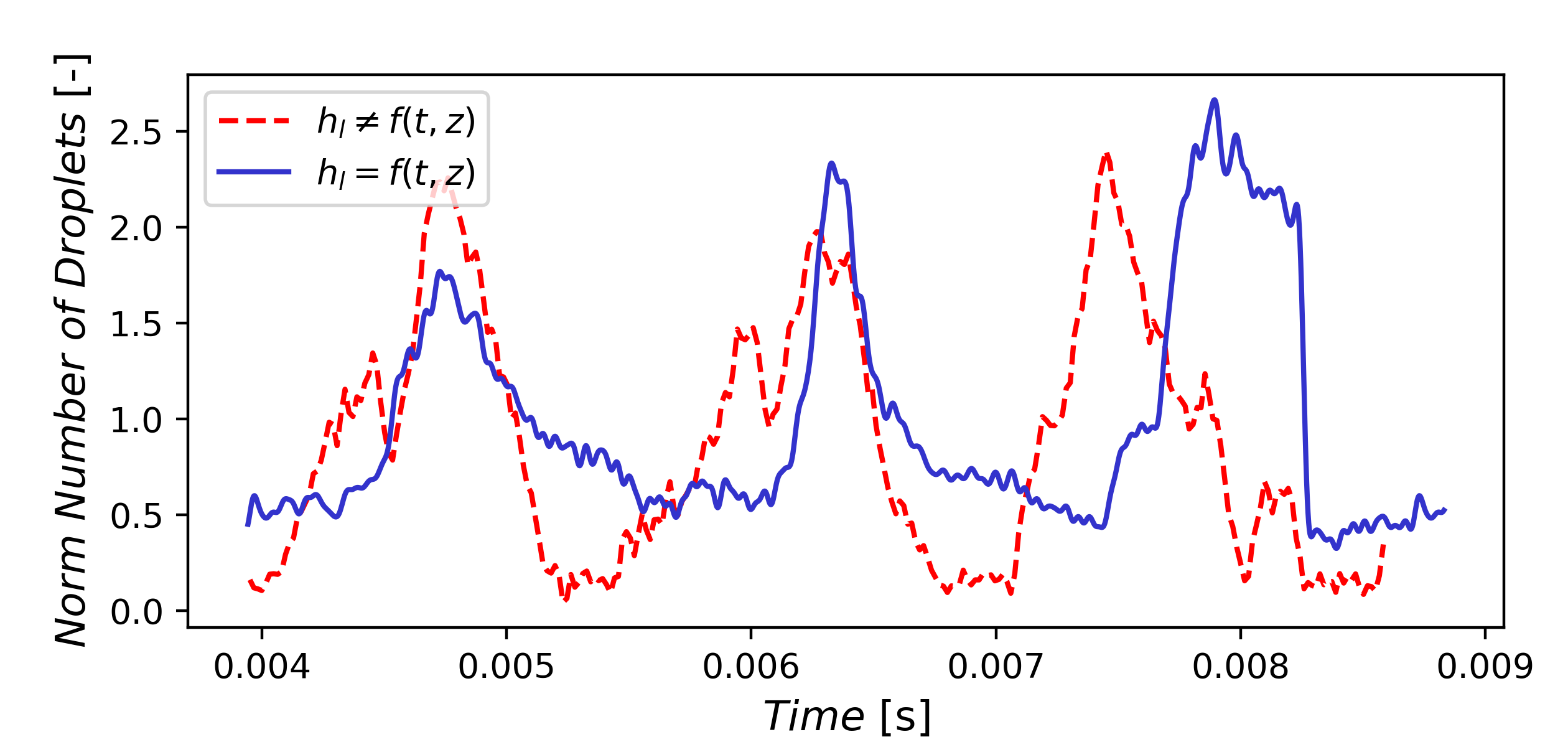}
%	\centering
	\caption{Temporal evolution of the number of detected droplets for both simulated cases. Results are normalized with the mean number of droplets detected in the time series in each case: 406 droplets for the $h_l \neq f(t,z)$, 62 droplets for the $h_l = f(t,z)$ case.}
	\label{fig:pospro_drops_vs_t}
\end{figure}

The post-processing technique described in Section \ref{sec:pospro} allowed detecting distinct droplets in both simulated cases. The temporal evolution of the number of detected droplets in each simulation is shown in Figure \ref{fig:pospro_drops_vs_t}, where the initial transient has been omitted and results are normalized with the mean amount of droplets detected in the depicted interval. These values are 406 droplets/frame in the $h_l \neq f(t,z)$ case and 62 droplets/frame in the $h_l = f(t,z)$ case. These large variation could be appreciated from the flow evolution depicted in Figure \ref{fig:results_DNS_qualitative}. While other literature works using VOF on the test case did not specifically report this quantity, their qualitative comparisons among experimental and computational snapshots also showed a clear overprediction of the amount of droplets computationally simulated (especially noticeable in the side views, considering the actual prefilmer span $b$ is an order of magnitude greater than the simulated span) \cite{Warncke2017, Mukundan2022}. This fact is aligned with the finding of the present investigation, in the sense that accounting for the liquid film evolution upstream of the prefilmer edge results in a less violent breakup and a substantially lower amount of generated droplets.

Coming back to Figure \ref{fig:pospro_drops_vs_t}, it is possible to distinguish three main breakup events where the number of droplets suddenly increases, substantially rising above the average. These instants correspond to the stages of \textit{bag breakup} illustrated in Figure \ref{fig:results_DNS_qualitative}. After a main breakup event, the number of generated droplets decreases as only ligament breakup is present while liquid is being accumulated in the reservoir behind the prefilmer edge prior to a new bag breakup event. Important differences are observed among the simulated cases. In the $h_l \neq f(t,z)$ case, the amount of generated droplets after a main event decays practically down to zero, importantly inhibiting atomization. In the $h_l = f(t,z)$ case, however, the amount of generated droplets only gets down to half of the average amount. This implies that ligament breakup is more relevant in this case and atomization is more continuous. As discussed in the view of Figure \ref{fig:pospro_drops_vs_t}, the fact that temporal and spanwise variations in the film thickness are included at the DNS inlet results in film waves reaching the liquid reservoir behind the perfilmer edge in a non-synchronized manner. Hence, the film reservoir is not uniform spanwise and bags are not generated along the whole span by the time some of them break. This yields a less violent main bag breakup event and a greater relative importance of the ligament breakup, as the resulting ligaments are also shaped by the remaining reservoir prior to the formation of new bags.

Additionally, the frequency among main events also differs for both simulations. A value of $f_{main} = \SI{656.2}{\hertz}$ is observed for the $h_l \neq f(t,z)$ case, whereas $f_{main} = \SI{630.0}{\hertz}$ is found for the $h_l = f(t,z)$ case. Even though both values are similar to the reported $\overline{f_{film}} = \SI{585.9}{\hertz}$ from the 2-phase flow precursor LES (recall Table \ref{tab:results_2phaseLES}), the $h_l = f(t,z)$ case is closer. Holz et al. \cite{Holz2018} already reported a direct relation among film wave frequency and breakup frequency, but justifying the minor differences through the slight decoupling by the accumulation of liquid at the trailing edge. This also justifies the differences observed in the present study among film wave frequency and breakup frequency. Even though the decoupling is present, not the same breakup frequency is retrieved if the film wave behavior is not introduced to the DNS. While this decoupling among frequencies is observed for this operating condition with $M = 15.7$, it must be noted that operating conditions with higher momentum flux ratios \cite{Fernandez2009} or with a lower ratio among film thickness and prefilmer thickness \cite{Inamura2019} (as is the case of commonly used aero engines) are expected to become dominated by the film evolution rather than by liquid accumulation. In such cases, the presented results highlight the importance of accounting for the liquid film evolution upstream of the prefilmer edge, either through the inlet boundary condition proposed in the study, or by means of a larger computational domain hardly affordable currently.

\begin{figure}[htbp]
	\centering
	\includegraphics[width=0.5\textwidth]{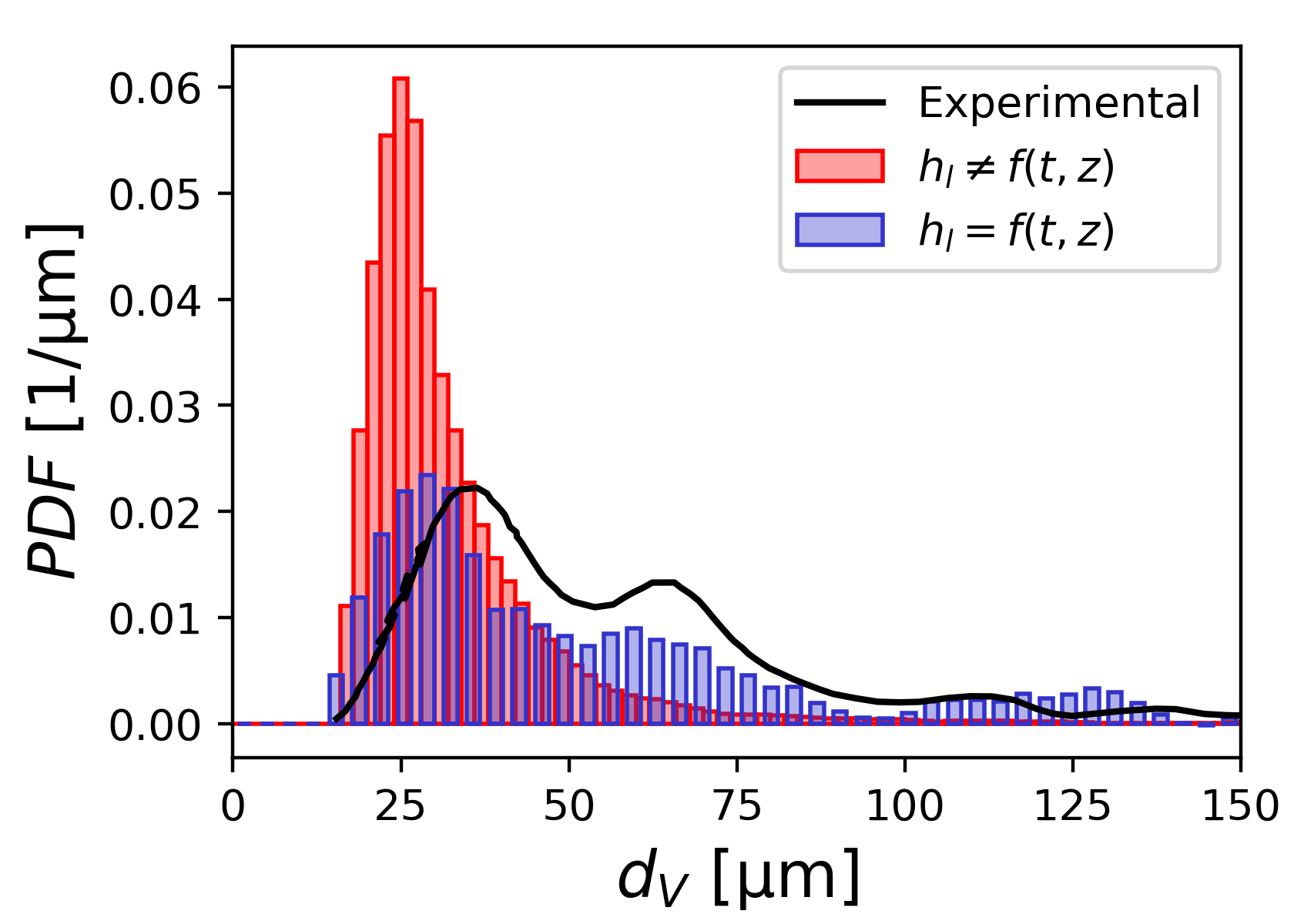}
	\caption{Droplet size Probability Density Function (PDF) for both simulations and experimental data from Warncke et al. \cite{Warncke2017}.}
	\label{fig:results_drops_diameters}
\end{figure}

As far as the droplet sizes are concerned, Figure \ref{fig:results_drops_diameters} shows the drop diameter PDFs for both simulated cases compared against experimental data. As mentioned in Section \ref{sec:pospro}, it is important to note that droplets exhibiting a diameter $d_v \leq 20$ have been discarded from the analysis, also corresponding to the resolution limit of the experimental data \cite{Warncke2017}. The comparison highlights that a substantially fairer agreement with experimental data is achieved when accounting for the liquid film history in the $h_l = f(t,z)$ case. The multi-modal shape is recovered in this case, correctly balancing the relative importance among \textit{bag breakup} and \textit{ligament breakup}, with a slight underprediction of the size values of the peaks in the first two modes. The $h_l \neq f(t,z)$ case did not exhibit this behavior. In this case, the ligament breakup mechanism is obscured by the large amount of droplets produced at the bag burst stage, which possess smaller sizes.

\begin{figure}[htbp]
	\centering
	\includegraphics[width=0.42\textwidth]{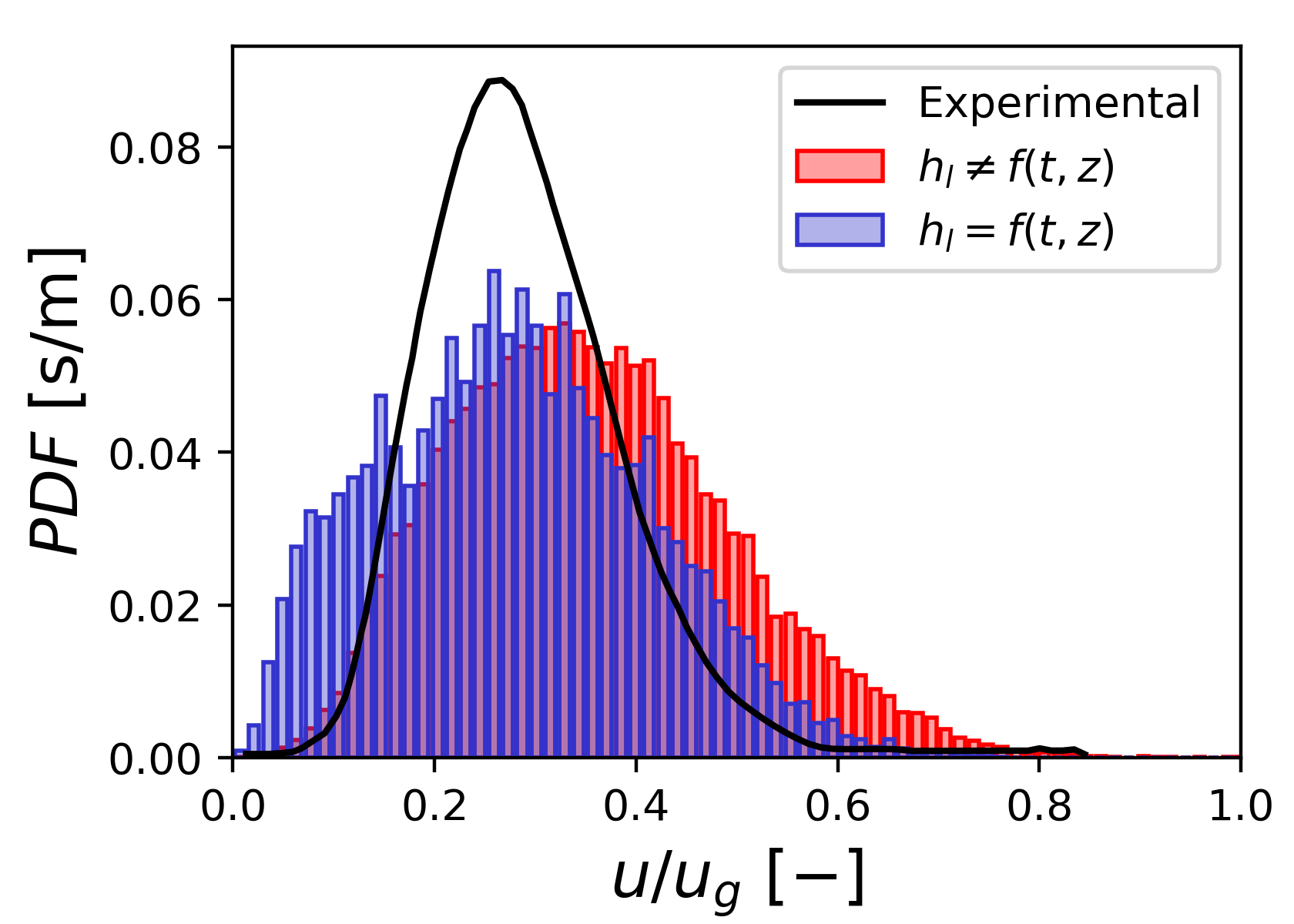}
	\includegraphics[width=0.42\textwidth]{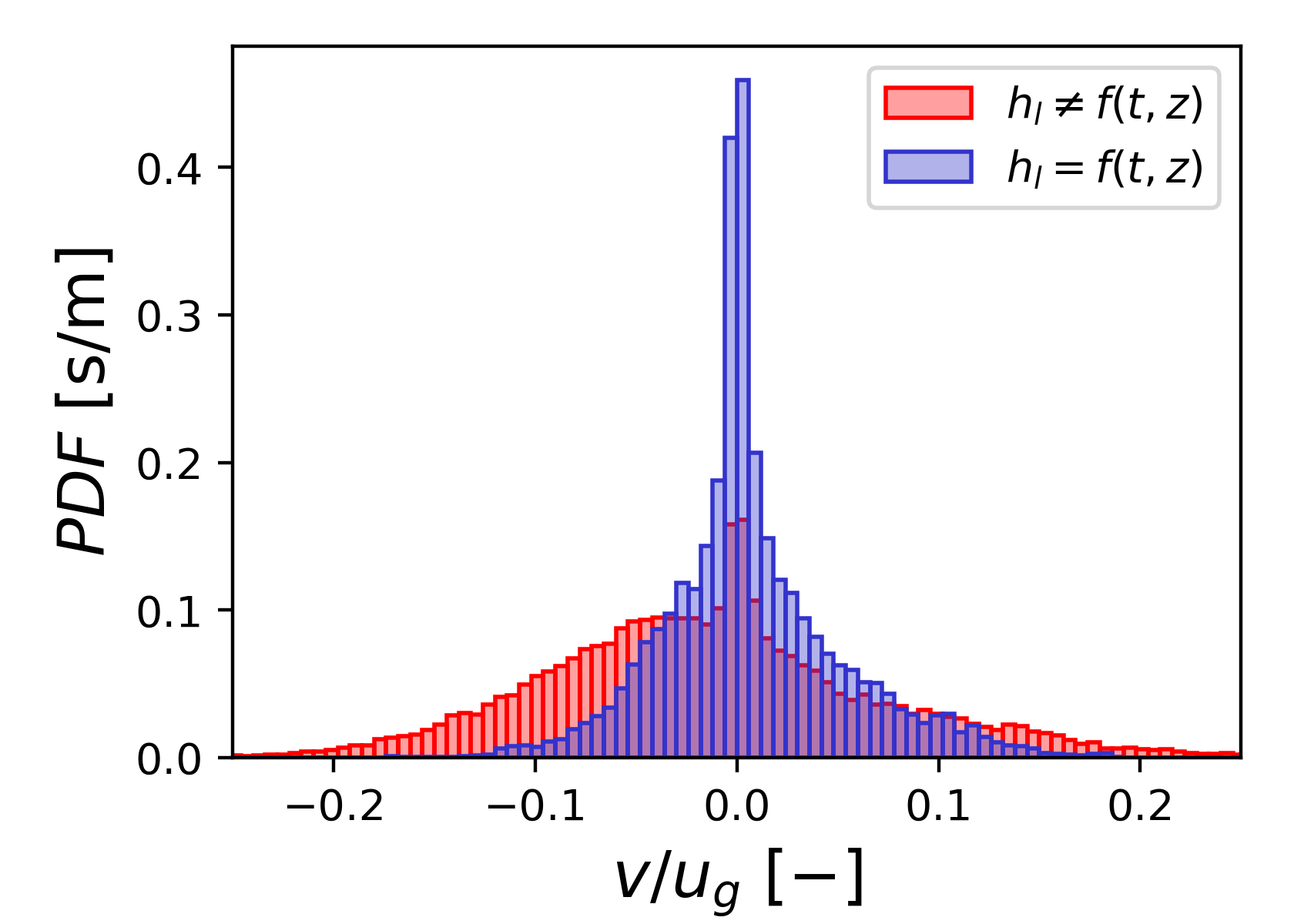}
	\includegraphics[width=0.42\textwidth]{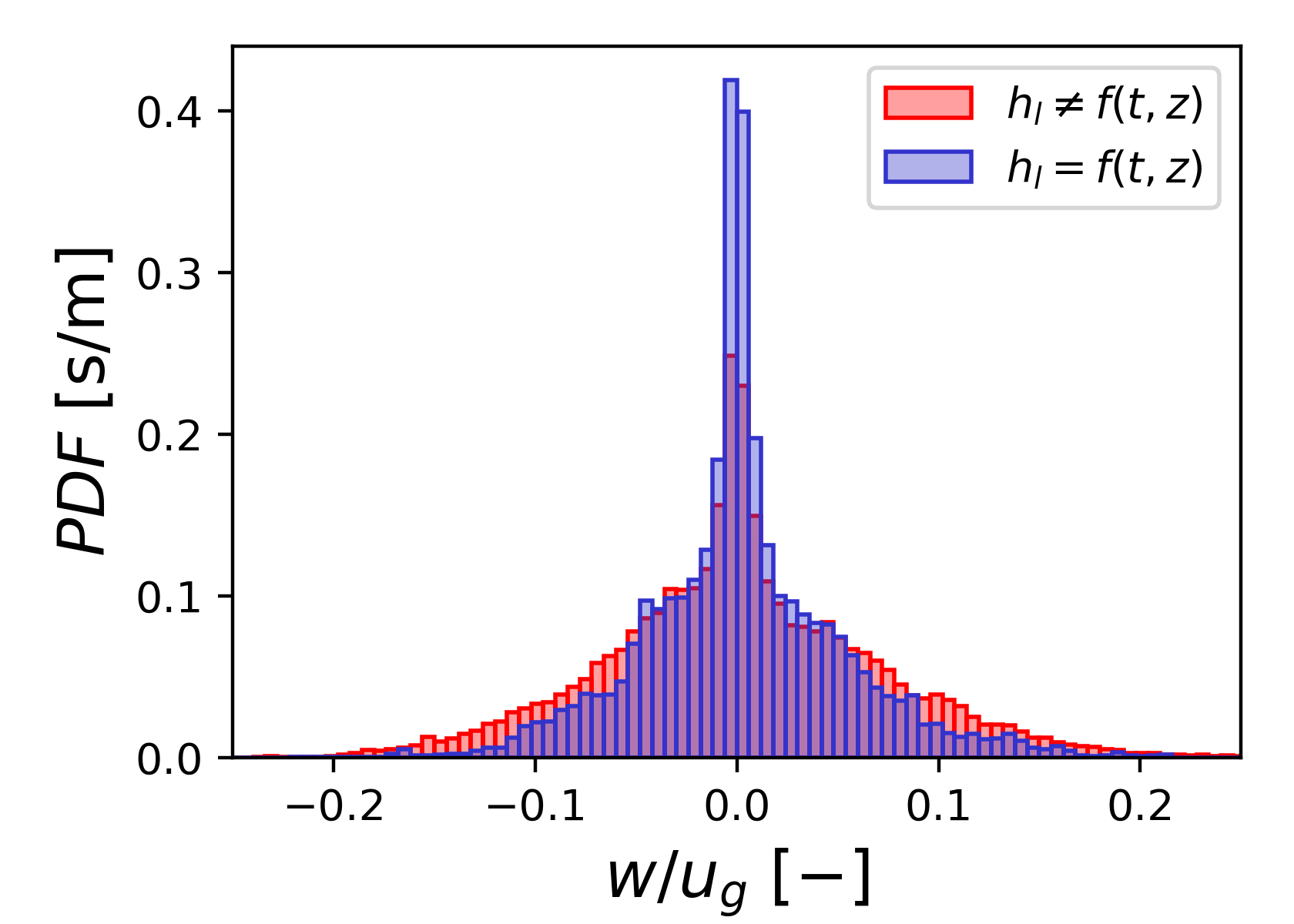}
%	\centering
	\caption{PDFs for the droplet velocity components for both simulated test cases: streamwise velocity (top left, including also experimental data from Warncke et al. \cite{Warncke2017}), wall-normal velocity (top right), spanwise velocity (bottom). All values are normalized with the gas bulk velocity.}
	\label{fig:results_drops_velocities}
\end{figure}

\begin{figure}[htbp]
	\centering
	\includegraphics[width=0.45\textwidth]{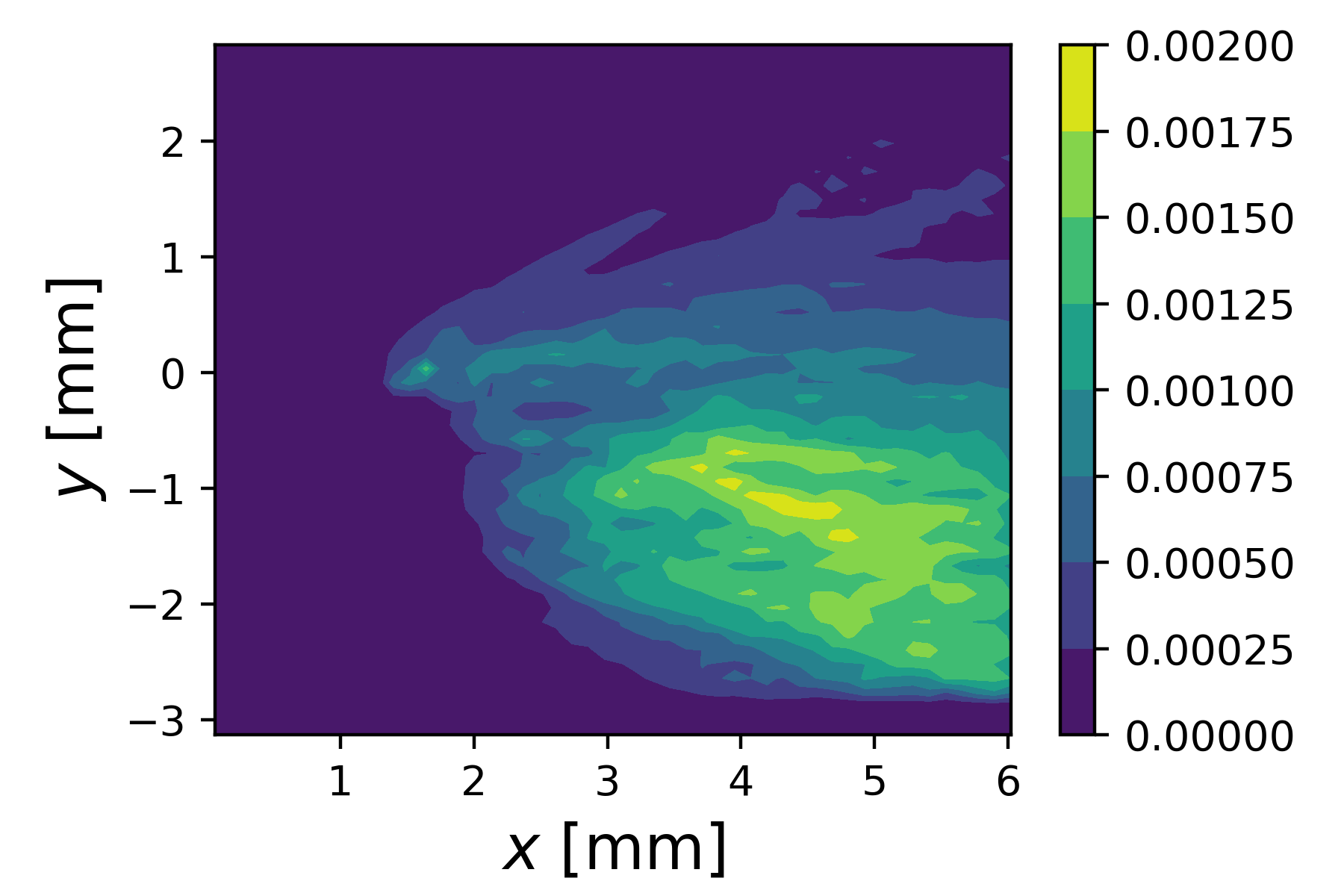}
	\includegraphics[width=0.45\textwidth]{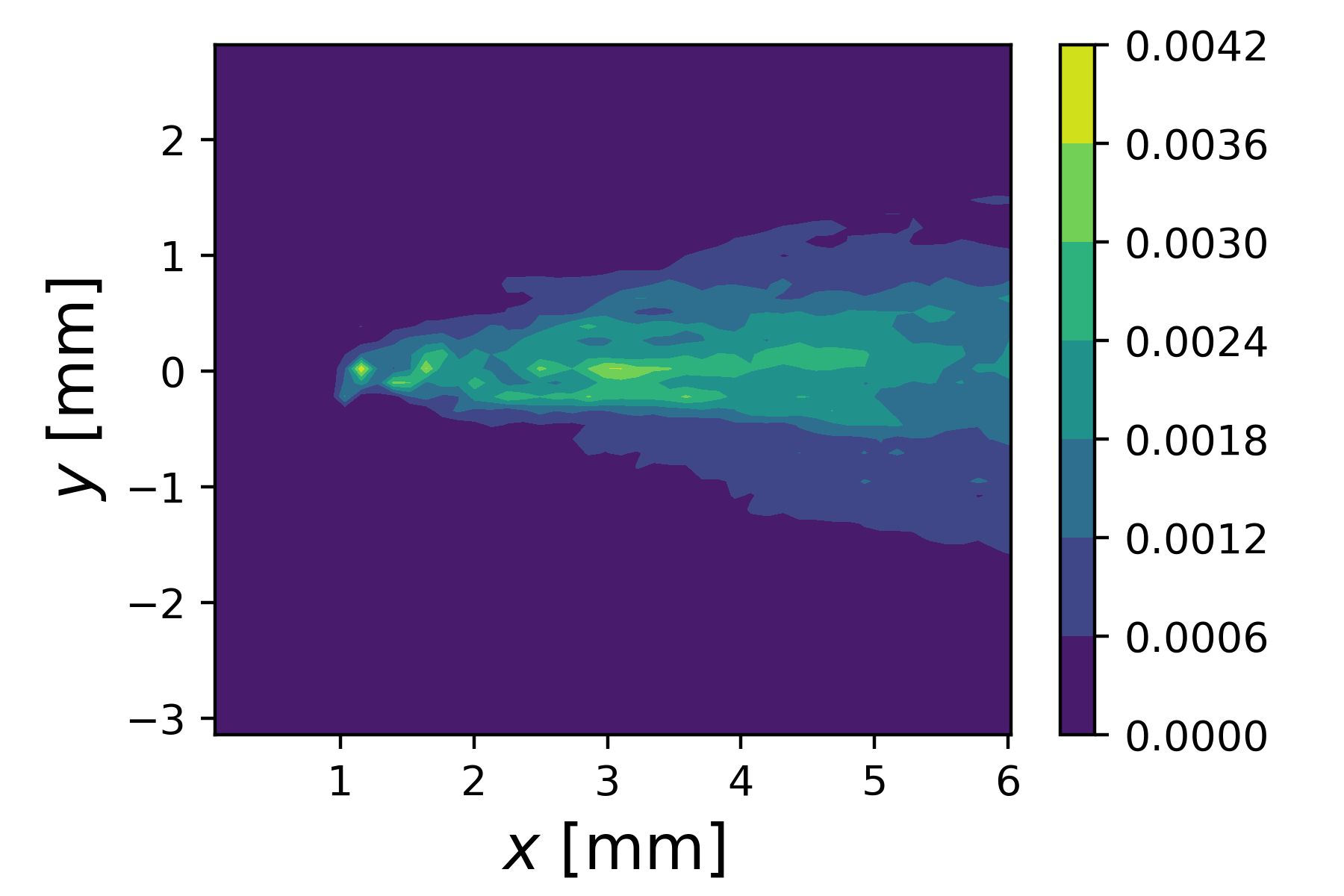}
%	\centering
	\caption{Contours displaying the probability of finding a droplet in the different locations of the XY plane: $h_l \neq f(t,z)$ case (left) and $h_l = f(t,z)$ case (right).}
	\label{fig:results_drops_position}
\end{figure}

Figure \ref{fig:results_drops_velocities} shows the PDFs obtained for each of the droplet velocity components, including a comparison with experimental data from \cite{Warncke2017} in the case of the streamwise component. Despite the differences in the number of predicted droplets and their size PDFs, the distributions are relatively similar among simulated cases. In the case of the streamwise velocity component, both predicted PDFs resemble a normal distribution. The agreement with experimental data seems slightly better in the $h_l = f(t,z)$ case, although values are underpredicted for a relevant amount of droplets. The distribution shape and dispersion about the average is similar for both simulated cases, but the distribution is shifted towards larger velocities in the $h_l \neq f(t,z)$ case. This fact is reasonable considering that a larger proportion of small droplets (subject to less aerodynamic drag when interacting with the gaseous phase) was found, and results in a substantial overprediction compared to experimental data. As far as the wall-normal velocities are concerned, the PDF is centered around 0 for the $h_l = f(t,z)$ case, exhibiting low dispersion. Nevertheless, the $h_l \neq f(t,z)$ case predicts a noticeably higher dispersion and its average is displaced towards negative values. As it happened for the ligament tips, this may be explained considering the locations at which it is more probable to find a droplet in each simulated case, as shown in Figure \ref{fig:results_drops_position}. As already explained in the view of Figure \ref{fig:results_DNS_qualitative}, in the $h_l \neq f(t,z)$ case the liquid film flaps more importantly about the prefilmer center line, resulting in a wider spray angle. Figure \ref{fig:results_drops_position} also highlights that droplets in the $h_l \neq f(t,z)$ case are preferentially found below the prefilmer center line, whereas they are quite uniformly distributed on both sides of the prefilmer in the $h_l = f(t,z)$ case. It is important to recall that, in the former case, ligaments were preferentially found above the prefilmer, as shown in Section \ref{subsec:results_ligs}. Figure \ref{fig:results_DNS_qualitative} showed a situation for this case for which ligaments indeed extended above the prefilmer. Such ligaments were generated from a film flapping motion whose bags got punctured towards the opposite side of the prefilmer, yielding a substantially higher amount of droplets below the prefilmer center line. Droplets generated below the prefilmer through this mechanism tend to travel further away from the prefilmer center line, reaching negative wall-normal velocities as reflected in Figure \ref{fig:results_drops_velocities}. In any case, it must be again noted that this shift towards negative wall-normal velocities may be biased by the fact that only 3 breakup events could be simulated. Results for the $h_l = f(t,z)$ case seem more statistically representative as the film accumulation behind the prefilmer edge is desynchronized spanwise, providing bag breakup both above and below the prefilmer center line and resulting in a spray more uniformly spread in the wall-normal direction. Finally, the differences among simulated cases in the spanwise component distributions are not so relevant, being centered around $w/u_g = 0$ in both cases and slightly more deviated about this value in the $h_l \neq f(t,z)$ case.

\begin{figure}[htbp]
	\centering
	\includegraphics[width=0.45\textwidth]{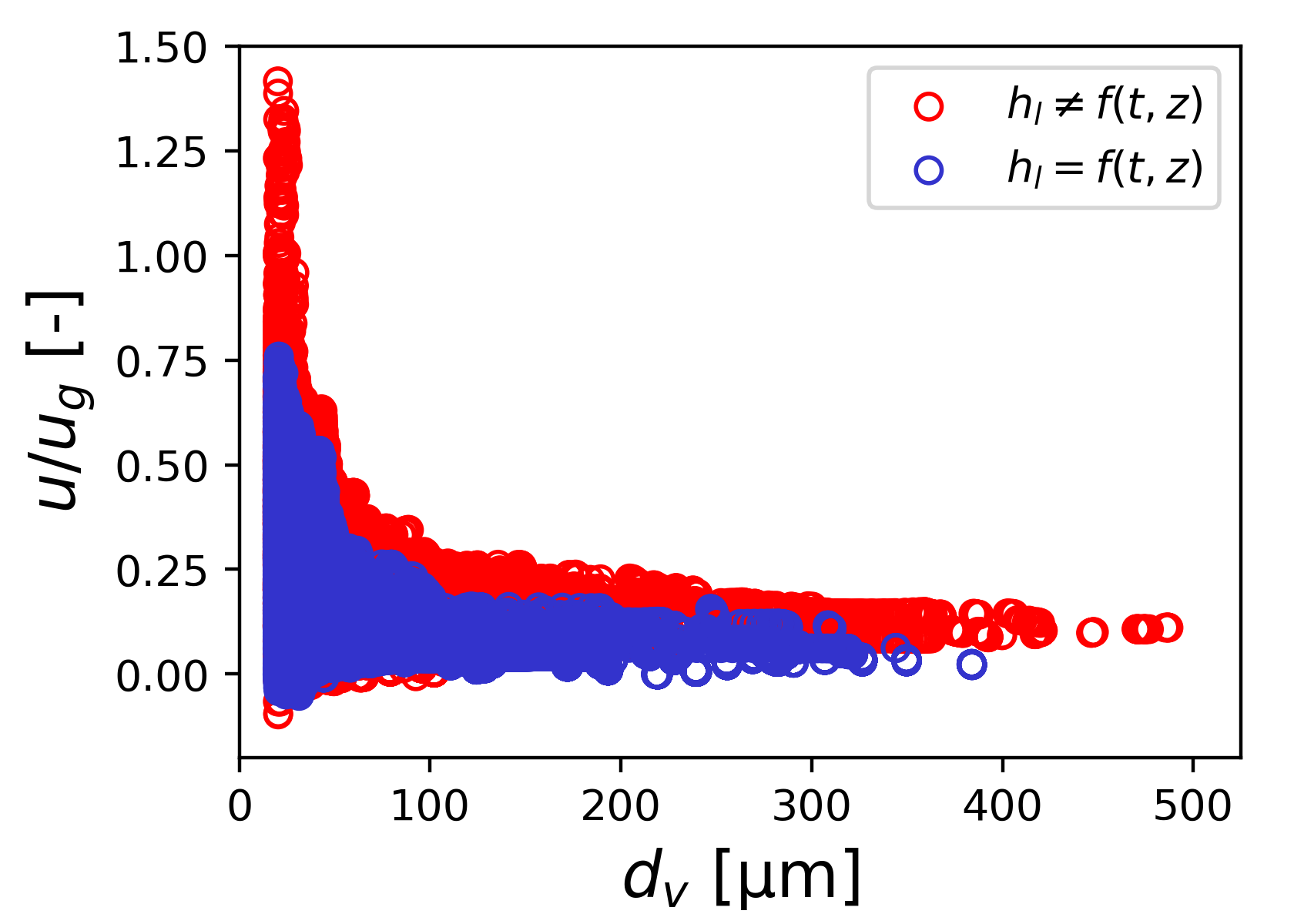}
	\includegraphics[width=0.45\textwidth]{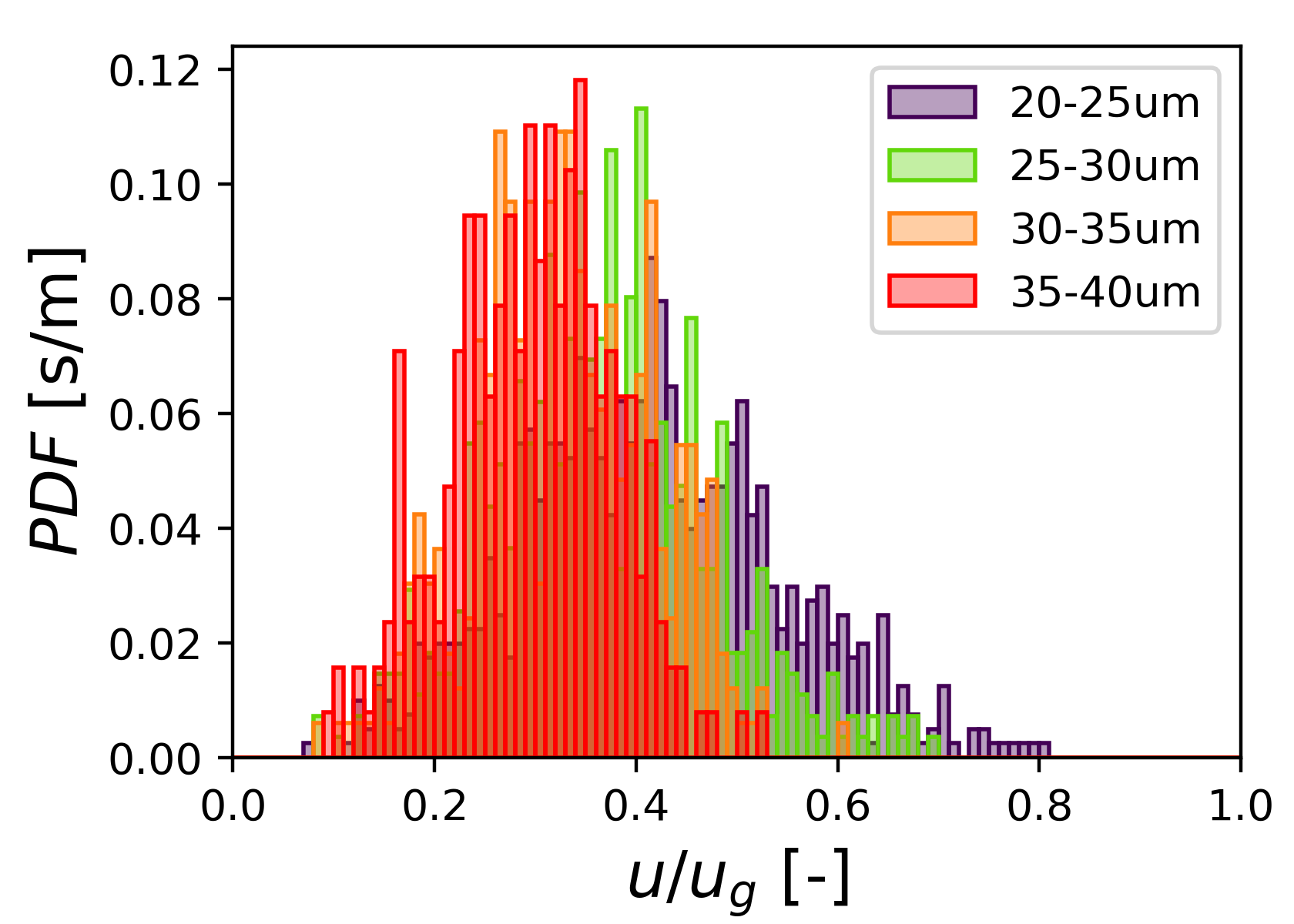}
	\caption{Scatter plot for the streamwise droplet velocity as a function of the drop diameter for both simulated cases (left), and detail on the streamwise droplet velocity PDFs corresponding to different drop size classes in the $h_l \neq f(t,z)$ case (right).}
	\label{fig:dropletsPDF_D70_ctevsvar_scatter_udv}
\end{figure}

It is interesting to analyze the correlation among droplet size and droplet velocity. Figure \ref{fig:dropletsPDF_D70_ctevsvar_scatter_udv} (left) shows the scattered information (focused on streamwise velocity) for all the detected droplets in both simulated cases. Results show that in both simulated cases large droplets statistically possess a lower streamwise velocity. Additionally, the dispersion in the velocity PDF is greater the lower the drop size, as confirmed in the view of Figure \ref{fig:dropletsPDF_D70_ctevsvar_scatter_udv} (right) where it is seen that the mean velocities are noticeably shifter towards high values as the droplet size decreases. These trends follow a non-linear fashion. As already implied by Figure \ref{fig:results_drops_velocities}, the comparison among test cases highlights the greater velocities achieved by the droplet cloud in the $h_l \neq f(t,z)$ case compared to the $h_l = f(t,z)$ case.

\section{Summary and Conclusions} \label{sec:conclusions}

In the present work, a numerical investigation on planar prefilming airblast primary breakup was conducted through a VOF-DNS method. A novel approach was presented to account for the spatio-temporal evolution of the liquid film thickness upstream of the prefilmer edge. This methodology required the computation of subsequent single-phase and two-phase flow LES to gather data to include at the DNS inflow boundary condition, considering the DNS domain had to be restricted to the last part of the prefilmer and the first millimetres downstream of its edge. These precursor simulations were successfully validated against consolidated channel flow DNS data and experimental data, respectively. Additionally, a methodology to post-process ligament data not only in a 2D projection but in a 3D domain was also proposed.

Results for an operating condition widely studied in the literature were discussed comparing two DNS simulations with different inflow conditions: a DNS that only accounted for a constant (timewise and spanwise) liquid film thickness at the domain inlet on the one hand; and the DNS that took advantage of the full proposed methodology by accounting for the oscillations in the liquid film thickness at the domain inlet on the other hand. The inflow condition of both simulations accounted for gas turbulence.

Qualitative comparisons among the primary breakup simulations showed important differences among simulation strategies. Even though the predicted breakup mechanism sequence was the same in accordance with the momentum flux ratio of the chosen operating condition (liquid accumulation behind the prefilmer edge followed by bag formation, bag breakup, ligament formation and ligament breakup), the resulting ligament distribution and droplet cloud were fairly different. Accounting for the liquid film evolution upstream of the prefilmer edge resulted in the liquid reservoir formed behind its edge being less uniformly distributed spanwise than it was for the constant liquid film thickness case, as the film crests and valleys do not reach the reservoir in a synchronized manner along the prefilmer span. This eventully results in a less violent main bag breakup event that generates a substantially lower amount of droplets from bag breakup and a proportionally higher amount of droplets from ligament breakup. The atomization thus seems inhibited to a certain extent, but is more continuous in time than the one found for the consntant liquid film thickness case.

DNS ligament and droplet cloud processed data were quantitatively validated against experimental results available in the literature. A reasonable agreement was found in both cases, with a noticeable improvement in the predictions when accounting for the liquid film evolution upstream of the prefilmer edge. The multi-modal feature of the droplet size distribution exhibited by the experiments could only be retrieved when making use of the proposed methodology. The observed differences in the velocity distributions, splitted among velocity components, could be explained by the fact that the spray was differently spread in the wall-normal direction in both cases. Accounting for the liquid film evolution along the prefilmer surface resulted in a uniformly spread spray within a relatively narrow cone. However, the constant liquid film thickness case resulted in a wider spray due to the greater importance achieved by film flapping in this case. In this simulation, the film behaves similarly at all spanwise locations for a given breakup event, meaning that all the shaped bags get punctured and breakup at the same stage of flapping, generating all droplets at the same side of the prefilmer, thereby also conditioning their velocities. In any case, the averaged and aggregated quantities considered in the analysis of the constant liquid film thickness case might be biased by the fact that only a few distinct breakup events were simulated and not accounting for a wide range of variability of breakup events. This limitation is in fact attenuated if the liquid film evolution along the prefilmer surface is accounted for.

Additionally to the commented shortcoming concerning the realizable simulated time, other limitations of the methodology must be highlighted. The large amount of resources required for the calculations implied only a small computational domain could be considered. This fact hindered the characterization of the ligament formation stage, as the experiments reported longer ligaments than the ones possibly found in the simulations. Nevertheless, this limitation is acknowledged in most literature works achieving a similar resolution than the one here presented. Precisely the mesh resolution is another limitation. Even though a proper turbulence reproduction was ensured by comparison with the Kolmogorov length scale, achieving lower cell sizes would have resulted in a better description of the liquid-gas interface, as a large amount of droplets was found to have diameters in the order of twice the cell size. 

Finally, it must be noted that the study is restricted to a given operating condition corresponding to the torn sheet breakup regime, for which liquid accumulation behind the prefilmer edge is significant and the breakup frequency is partially decoupled from the film wave frequency. Conditions with a higher gas-to-liquid momentum flux ratio, realizable in aero engines, lead to the so-called membrane sheet breakup, where the accumulation mechanism loses importance in favor of a direct disintegration in ligaments and droplets behind the prefilmer edge. In such circumstances the breakup frequency is expected to be more closely coupled to the film wave frequency. This should result in a more realistic description and more accurate predictions being obtained by the methodology proposed in the present investigation.

\section*{Acknowledgements}
Research leading to these results has received funding from the Clean Sky 2 Joint Undertaking European Union's Horizon 2020 research and innovation programme through the ESTiMatE project, grant agreement 821418. The authors acknowledge PRACE for awarding access to computational resources on JOLIOT CURIE-AMD at GENCI@CEA, France (proposal 2019204944). Computer resources at Marenostrum Supercomputer and the technical support provided by Barcelona Supercomputing Center (RES IM-2020-3-0018 and IM-2021-1-0010) in the frame of the Spanish Supercomputing Network is also thankfully acknowledged. Additionally, the support given to Mr. Carlos Moreno by Universitat Polit\`ecnica de Val\`encia through the "FPI Subprograma 2" grant within the "Programa de Apoyo para la Investigaci\'on y Desarrollo (PAID-01-19)" is acknowledged. The authors must also thank Marco Crialesi and Wojciech Aniszewski for their technical advice on custom modifications on PARIS, and Johan Sundin for his implementation of the contact angle model. Help from Achille Schmitter, Hugo Mart\'inez and Lucas Gonz\'alez processing the single-phase LES, two-phase LES, and DNS, respectively, is also appreciated.

\bibliography{export}

\end{document}